\documentclass[prd,nofootinbib,showpacs,preprint]{revtex4}
\usepackage{amsmath}
\usepackage{graphicx}
\usepackage{epsfig}
\graphicspath{{Figs/}}
\usepackage{dcolumn}
\usepackage{bm}
\usepackage{amssymb}
\usepackage[usenames,dvipsnames]{color}
\usepackage{slashed}
\usepackage[dvipdfm,colorlinks,citecolor=blue]{hyperref}
\begin{document}
\title{Perturbations to $\mu-\tau$ Symmetry, Leptogenesis and  \\Lepton Flavour Violation with Type II Seesaw}
\author{Manikanta Borah}
\email{mani@tezu.ernet.in}
\author{Debasish Borah}
\email{dborah@tezu.ernet.in}
\author{Mrinal Kumar Das}
\email{mkdas@tezu.ernet.in}
\affiliation{Department of Physics, Tezpur University, Tezpur-784028, India}
\author{Sudhanwa Patra}
\email{sudha.astro@gmail.com}
 \affiliation{Center of Excellence in Theoretical and Mathematical Sciences, 
Siksha O Anusandhan University, Bhubaneswar-751030, India}
\begin{abstract}
We study the possibility of generating non-zero reactor mixing angle $\theta_{13}$ by perturbing the $\mu-\tau$ symmetric neutrino mass matrix. 
The leading order $\mu-\tau$ symmetric neutrino mass matrix originates from type I seesaw mechanism whereas the perturbations to $\mu-\tau$ 
symmetry originate from type II seesaw term. We consider four different realizations of $\mu-\tau$ symmetry: Bimaximal Mixing(BM), Tri-bimaximal 
Mixing (TBM), Hexagonal Mixing (HM) and Golden Ratio Mixing (GRM) all giving rise to $\theta_{13} = 0, \theta_{23} = \frac{\pi}{4}$ but different 
non-zero values of solar mixing angle $\theta_{12}$. We assume a minimal $\mu-\tau$ symmetry breaking type II seesaw mass matrix as a perturbation 
and calculate the neutrino oscillation parameters as a function of type II seesaw strength. We then consider the origin of non-trivial leptonic 
CP phase in the charged lepton sector and calculate the lepton asymmetry arising from the lightest right handed neutrino decay by incorporating 
the presence of both type I and type II seesaw. We constrain the type II seesaw strength as well as leptonic CP phase (and hence the charged 
lepton sector) by comparing our results with experimental neutrino oscillation parameters as well as Planck bound on baryon to photon ratio. 
Finally, we extend our analysis on lepton flavour violating decays like $\mu \to e \gamma$ and 
$\mu \to eee$ due to exchange of TeV scale Higgs triplet scalar within the low scale type II seesaw framework. The branching ratios for these 
lepton flavour processes are examined with the small type II perturbation term $\omega$ and the estimated values are very close to the 
experimental bound coming from current search experiments.
\end{abstract}
\pacs{12.60.-i,12.60.Cn,14.60.Pq}
\maketitle
\section{Introduction}
The standard model (SM) of particle physics have been established as the most successful theory describing all fundamental 
particles and their interactions except gravity, specially after the discovery of its last missing piece, the Higgs boson 
in 2012. In spite of its huge phenomenological success, the SM fails to explain many observed phenomena in nature. Origin 
of tiny neutrino mass and matter-antimatter asymmetry are two of such phenomena which can be explained only within the framework 
of some beyond standard model (BSM) physics. Neutrinos which remain massless in the SM, have been shown to have tiny but non-zero 
mass (twelve order of magnitude smaller than the electroweak scale) by several neutrino oscillation experiments \cite{PDG}. 
Recent neutrino oscillation experiments T2K \cite{T2K}, Double ChooZ \cite{chooz}, Daya-Bay \cite{daya} and RENO \cite{reno} 
have not only made the earlier predictions for neutrino parameters more precise, but also predicted non-zero value of the 
reactor mixing angle $\theta_{13}$ as given below
\begin{table}[h!]
\begin{tabular}{ccc}
        \hline 
Experimental Data          & $\sin^2 2\theta_{13}$        
                                                    &  $\sin^2 \theta_{13}$    \\
        \hline \hline
T2K \cite{T2K}             & $0.11^{0.11}_{-0.05}\, \left(0.14^{0.12}_{-0.06} \right)$ 
                                                    &  $0.028^{0.019}_{-0.024}\, \left(0.036^{0.022}_{-0.030} \right)$ \\
Double ChooZ \cite{chooz}  & $0.086 \pm 0.041 \pm 0.030$ 
                                                    &  $0.022^{0.019}_{-0.018}$ \\
Daya-Bay \cite{daya}       & $0.092 \pm 0.016 \pm 0.005$ 
                                                    &  $0.024 \pm 0.005$ \\
RENO \cite{reno}           & $0.113 \pm 0.013 \pm 0.019$ 
                                                    &  $0.029 \pm 0.006$ \\
                                                    \hline
\end{tabular}
\caption{Experimental value of reactor mixing angle from recent neutrino oscillation experiments.}
\label{table-theta13}
\end{table}
Recent global fits for different oscillation parameters within their $3\sigma$ range taken from Ref.\cite{schwetz12} and 
Ref.\cite{global0} are presented below in table.\ref{table-osc}.

Several BSM frameworks have been proposed to explain the origin of tiny neutrino mass and the pattern of neutrino mixing. 
Tiny neutrino mass can be explained by seesaw mechanisms which broadly fall into three types : type I \cite{ti}, type II \cite{tii} 
and type III \cite{tiii} whereas the pattern of neutrino mixing can be understood by incorporating additional flavor symmetries.

\begin{table}[t]
\begin{tabular}{ccc}
        \hline 
Oscilation Parameters  &  Within 3$\sigma$ range         &  within 3$\sigma$ range   \\
           &   ({\it Schwetz et al.}\cite{schwetz12})   &   ({\it Fogli et al.}\cite{global0}) \\
        \hline \hline
$\Delta m^2_{\rm {21}} [10^{-5} \mbox{eV}^2]$              & 7.00-8.09   & 6.99-8.18     \\
$|\Delta m^2_{\rm {31}}(\mbox{NH})| [10^{-3} \mbox{eV}^2]$ & 2.27-2.69   & 2.19-2.62     \\
$|\Delta m^2_{\rm {23}}(\mbox{IH})| [10^{-3} \mbox{eV}^2]$ & 2.24-2.65   & 2.17-2.61     \\
\hline
$\sin^2\theta_{12}$                                        & 0.27-0.34   & 0.259-0.359   \\
$\sin^2\theta_{23}$                                        & 0.34-0.67   & 0.331-0.637   \\
$\sin^2\theta_{13}$                                        & 0.016-0.030 & 0.017-0.031  \\
        \hline
\end{tabular}
\caption{The global fit values for the mass squared differences and mixing angles as reported 
         by Ref.\cite{schwetz12} presented by 2nd column and by Ref.\cite{global0} by third column.}
\label{table-osc}
\end{table}

The neutrino oscillation data before the discovery of non-zero $\theta_{13}$ were in perfect agreement with $\mu-\tau$ symmetric neutrino mass matrix. Four different neutrino mixing pattern which can originate from such a $\mu-\tau$ symmetric neutrino mass matrix are: Bimaximal Mixing (BM) \cite{bimax}, Tri-bimaximal Mixing (TBM) \cite{Harrison}, Hexagonal Mixing (HM) \cite{HexM} and Golden Ratio Mixing (GRM) \cite{GoldenM}. All these scenarios predict $\theta_{23} = 45^o, \theta_{13} = 0$ but different values of solar mixing angle $\theta_{12} = 45^o$ (BM), $\theta_{12}=35.3^o$ (TBM), $\theta_{12}=30^o$ (HM), $\theta_{12}=31.71^o$ (GRM). However, in view of the fact that the latest experimental data have ruled out $\text{sin}^2\theta_{13}=0$, one needs to go beyond these $\mu-\tau$ symmetric frameworks. Since the experimental value of $\theta_{13}$ is still much smaller than the other two mixing angles, $\mu-\tau$ symmetry can still be a valid approximation and the non-zero $\theta_{13}$ can be accounted 
for by 
incorporating the presence of small perturbations to $\mu-\tau$ symmetry coming from different sources like charged lepton mass diagonalization, for example. Several such scenarios have been widely discussed in \cite{nzt13, nzt13GA} and the latest neutrino oscillation data can be successfully predicted within the framework of many interesting flavor symmetry models.

Apart from the origin of neutrino mass and mixing, the observed matter antimatter asymmetry also remains unexplained within the SM framework. The observed baryon asymmetry in the Universe is encoded in the baryon to photon ratio measured by dedicated cosmology experiments like Wilkinson Mass Anisotropy Probe (WMAP), Planck etc. The latest data available from Planck mission constrain the baryon to photon ratio \cite{Planck13} as
\begin{equation}
Y_B \simeq (6.065 \pm 0.090) \times 10^{-10}
\label{barasym}
\end{equation} 
Leptogenesis is one of the most widely studied mechanism of generating this observed baryon asymmetry in the Universe by generating an asymmetry in the leptonic sector first and later converting it into baryon asymmetry through electroweak sphaleron transitions \cite{sphaleron}. As pointed out first by Fukugita and Yanagida \cite{fukuyana}, the out of equilibrium CP violating decay of heavy Majorana neutrinos provides a natural way to create the required lepton asymmetry. The salient feature of this mechanism is the way it relates two of the most widely studied problems in particle physics: the origin of neutrino mass and the origin of matter-antimatter asymmetry. This idea has been implemented in several interesting models in the literature \cite{leptoreview,joshipura,davidsonPR}. Recently such a comparative study was done to understand the impact of mass hierarchies, Dirac and Majorana CP phases on the predictions for baryon asymmetry in \cite{leptodborah} within the framework of left-right symmetric 
models.

In the present work we propose a common mechanism which can generate the desired neutrino mass and mixing including non-zero $\theta_{13}$ and 
also the matter antimatter asymmetry. We extend the SM by three right handed singlet neutrinos and one Higgs triplet such that both type I and 
type II seesaw can contribute to neutrino mass. Type I seesaw is assumed to give rise to a $\mu-\tau$ symmetric neutrino mass matrix with 
$\theta_{13}=0$ whereas type II seesaw acts as a perturbation which breaks the $\mu-\tau$ symmetry resulting in non-zero $\theta_{13}$. 
Similar works have been done recently where type II seesaw was considered to be the origin of $\theta_{13}$ \cite{db-t2} as well as non-zero 
Dirac CP phase $\delta$ \cite{dbijmpa} by assuming the type I seesaw giving rise to TBM type mixing. Some earlier works studying neutrino masses 
and mixing by using the interplay of two different seesaw mechanisms can be found in \cite{devtbmt2, mkd-db-rm, dbgrav}. 
In this work we generalize earlier studies on TBM type mixing to most general $\mu-\tau$ symmetric neutrino mass matrices and 
check whether a minimal form of $\mu-\tau$ symmetry breaking type II seesaw can give rise to correct value of reactor mixing angle 
$\theta_{13}$. We then calculate the predictions for other neutrino parameters as well as observables like sum of absolute neutrino 
masses $\sum_i \lvert m_i \rvert$ and effective neutrino mass $m_{ee} = \lvert \sum_i U^2_{ei} m_i \rvert$. We check whether the sum 
of absolute neutrino masses obey the cosmological upper bound $\sum_i \lvert m_i \rvert < 0.23$ eV \cite{Planck13} and whether the 
effective neutrino mass $m_{ee}$ lies within the bounds coming from neutrinoless double beta decay experiments. We also calculate 
the lepton asymmetry by considering the source of leptonic Dirac CP violation in the charged lepton sector. From the requirement 
of generating correct neutrino parameters and baryon asymmetry we constrain type II seesaw strength, Dirac CP phase and at the 
same time discriminate between neutrino mass hierarchies, different lightest neutrino masses and different $\mu-\tau$ symmetric 
mass matrices.

To the end, the lepton flavour violating decays like $\mu \to e \gamma$ and $\mu \to 3e$ with mediation of TeV scale Higgs triplet 
scalar has been carefully examined. In the present work, we have considered type-II seesaw contribution to light neutrino mass $m_\nu = 
f_\nu v_\Delta$, where $v_\Delta \simeq \mu_{\Phi \Delta} v^2 {\Large /} \sqrt{2} M^2 $ being the induced VEV of the neutral 
component of the Higgs triplet scalar, as sub-dominant term (of the order of $0.001$ eV) as compared to dominant type I seesaw 
contribution. If $f_\nu$ is assumed to be taken its natural value {\cal O}(0.1-1.0), then sub-eV scale light neutrino mass can be 
generated by two fold way: (i) either by large Higgs triplet mass $M_{\Delta}$ \cite{tii} (ii) or by small value of $\mu_{\Phi \Delta}$. 
It makes the possibility of probing standard type II seesaw mechanism at high energy accelerator experiments with large seesaw 
scale $M_\Delta > 10^{13}$ GeV and the associated Lepton Flavour Violating (LFV) processes are heavily suppressed. Alternatively, 
if the type II seesaw mechanism is operative at TeV scale, then the resulting LFV processes are prominent and same-sign dilepton 
signatures is one of the prime focus at LHC. Within low scale type II set up, the seesaw relation for Higgs triplet VEV is consistent 
with small trilinear mass term $\mu_{\Phi \Delta}$ and TeV scale Higgs scalar triplet mass. We wish to examine the associated LFV 
processes which can be originated via Higg triplet scale and the corresponding branching ratios for them are significant enough 
to be probed at ongoing search experiments if the mass of the triplet scalar is in the TeV range.

The plan of the paper is sketched as following manner. In section \ref{sec:seesaw} we discuss the methodology of type I and type II 
seesaw mechanisms. In section \ref{sec:mutau}, we discuss the parametrization of different $\mu-\tau$ symmetric neutrino mass matrices. 
We then discuss deviations from $\mu-\tau$ symmetry using type II seesaw in section \ref{sec:devmutau}. In section \ref{sec:lepto}, 
we discuss CP violation and outline the mechanism of leptogenesis in the presence of type I and type II seesaw. In section \ref{sec:numeric} 
we discuss our numerical analysis and results and then finally conclude in section \ref{sec:conclude}.

\section{Seesaw Mechanism: Type I and Type II}
\label{sec:seesaw}
Type I seesaw \cite{ti} mechanism is the simplest possible realization of the dimension five Weinberg operator \cite{weinberg} 
for the origin of neutrino masses within a renormalizable framework. This mechanism is implemented in the standard model by the 
inclusion of three additional right handed neutrinos $(\nu^i_R, i = 1,2,3)$ as $SU(2)_L$ singlets with zero $U(1)_Y$ charges. 
Being singlet under the gauge group, bare mass terms of the right handed neutrinos $M_{RR}$ are allowed in the Lagrangian. 
On the other hand, in type II seesaw \cite{tii} mechanism, the standard model is extended by inclusion 
of an additional $SU(2)_L$ triplet scalar field $\Delta$ having $U(1)_Y$ charge twice that of lepton doublets with 
its $2 \times 2$ matrix representation as
\begin{equation}
\Delta =
\left(\begin{array}{cc}
\Delta^+/\sqrt{2}   &   \Delta^{++} \\
\Delta^0            &  -\Delta^+/\sqrt{2}
\end{array}\right) \nonumber \, .
\end{equation} 

Thus, the gauge invariant lagrangian relevant for type I plus type II seesaw mechanism is given below
\begin{eqnarray}
\mathcal{L}=(D_\mu \Phi)^\dagger (D^\mu \Phi) + \mbox{Tr}(D_\mu \Delta)^\dagger (D^\mu \Delta) -\mathcal{L}^{\rm lept}_Y - V(\Phi, \Delta)\, ,
\end{eqnarray}
with the leptonic interaction terms, 
\begin{eqnarray}
\mathcal{L}_Y = y_{ij} \ell_i \tilde{\Phi} \nu_R + f_{ij} \ell^T_{i} C (i \tau_2) \Delta \ell_j 
                + \frac{1}{2} \nu^T_R C^{-1} M_R \nu_R + \text{h.c.}
\end{eqnarray}
Here $\ell_L \equiv(\nu,~e)^T_L$, $\Phi \equiv (\phi^0 ,\phi^-)^T$ and C is the charge conjugation operator. 
The scalar potential of the model using SM Higgs doublet $\Phi$ and Higgs triplet scalar $\Delta_L$ is 
\begin{eqnarray}
\mathcal{V}(\Phi, \Delta) &=& \mu^2_{\Phi} \Phi^\dagger \Phi + \lambda_1 \left(\Phi^\dagger \Phi \right)^2  
           + \mu^2_{\Delta} \mbox{Tr}\left(\Delta^\dagger \Delta \right) 
           + \lambda_2 \left[\mbox{Tr}\left(\Delta^\dagger \Delta \right) \right]^2
\nonumber \\
         &+& \lambda_3 \mbox{Det}\left(\Delta^\dagger \Delta \right) 
          +\lambda_4 \left(\Phi^\dagger \Phi \right) \mbox{Tr}\left(\Delta^\dagger \Delta \right)
          +\lambda_5 \left(\Phi^\dagger \tau_i \Phi \right) \mbox{Tr}\left(\Delta^\dagger \tau_i \Delta \right) \nonumber \\
         &+& \frac{1}{\sqrt{2}} \mu_{\Phi \Delta} \left(\Phi^T i \tau_2 \Delta \Phi \right) + \text{h.c.}
         \label{eq:scalar_lag}
\end{eqnarray}

With  vacuum expectation value of the SM Higgs $\langle \Phi^0 \rangle =v/\sqrt{2}$, the trilinear mass term 
$\mu_{\Phi \Delta}$ generates an induced VEV for Higgs triplet as $\langle \Delta^0 \rangle=v_\Delta/\sqrt{2}$ where 
$v_\Delta \simeq \mu_{\Phi \Delta} v^2{\Large /} \sqrt{2} M^2 $, 
the resulting in $6\times 6$ neutrino mass matrix after electroweak symmetry breaking reads as
\begin{equation}
\mathcal{M}_\nu= \left( \begin{array}{cc}
              m_{LL} & m_{LR}   \\
              m^T_{LR} & M_{RR}
                      \end{array} \right) \, ,
\label{eqn:numatrix}       
\end{equation}
where $m_{LR}=y_\nu\,v$ is the Dirac neutrino mass, $m_{LL}=f_\nu \, v_\Delta$ is the Majorana mass for 
light active neutrinos and $m_{RR}$ is the bare mass term for heavy sterile Majorana neutrinos.
Within the mass hierarchy $M_{RR} \gg m_{LR} \gg m_{LL}$, the seesaw formula for light neutrino mass 
is given by
\begin{equation}
m_\nu\equiv m_{LL}=m_{LL}^{I} + m_{LL}^{II}
\label{type2a}
\end{equation}
where the formula for type I  seesaw contribution is presented below,
\begin{equation}
m_{LL}^I=-m_{LR} M_{RR}^{-1} m_{LR}^{T}.
\end{equation}
where $m_{LR}$ is the Dirac mass term of the neutrinos which is typically of electroweak scale. 
Demanding the light neutrinos to be of eV scale one needs $M_{RR}$ to be as high as $10^{14}$ GeV 
without any fine-tuning of Dirac Yukawa couplings. Whereas the type II seesaw contribution to light neutrino 
mass is given by
\begin{equation}
m_{LL}^{II}= f_\nu v_\Delta\, ,
\end{equation}
where the analytic formula for induced VEV for neutral component of the Higgs scalar triplet, derived from 
the minimization of the scalar potential, is
\begin{equation}
v_\Delta \equiv \langle \Delta^0 \rangle = \frac{\mu_{\Phi \Delta} v^2}{M^2_{\Delta}}\, .
\label{vev} 
\end{equation}
In the low scale type II seesaw mechanism operative at TeV scale, barring the naturalness issue, one can consider 
a very small value of trilinear mass parameter to be
$$\mu_{\Phi \Delta} \simeq 10^{-8}\, \mbox{GeV}\, ,$$
where the Higgs scalar triplet mass lie within TeV range which give interesting phenomenological possibility 
of being produced in pairs at LHC. The sub-eV scale light neutrino mass with type II seesaw mechanism contrains 
the corresponding Majorana Yukawa coupling as
$$f^2_\nu < 1.4 \times 10^{-5} \left(\frac{M_\Delta}{\mbox{1\, TeV}} \right)\, . $$
Within reasonable value of $f_\nu \simeq 10^{-2}$, the triplet Higgs scalar VEV is $v_\Delta \simeq 10^{-7}$ GeV 
which is in agreement with the oscillation data. It is worth to note here that the tiny trilinear mass 
parameter $\mu_{\Phi \Delta}$ controls the neutrino overall mass scale, but does not play any role in the couplings with 
the fermions and thereby, making the lepton flavour violion studies more viable.
%
\begin{figure}[htb!]
\centering
\includegraphics[width=0.82\textwidth,angle=0]{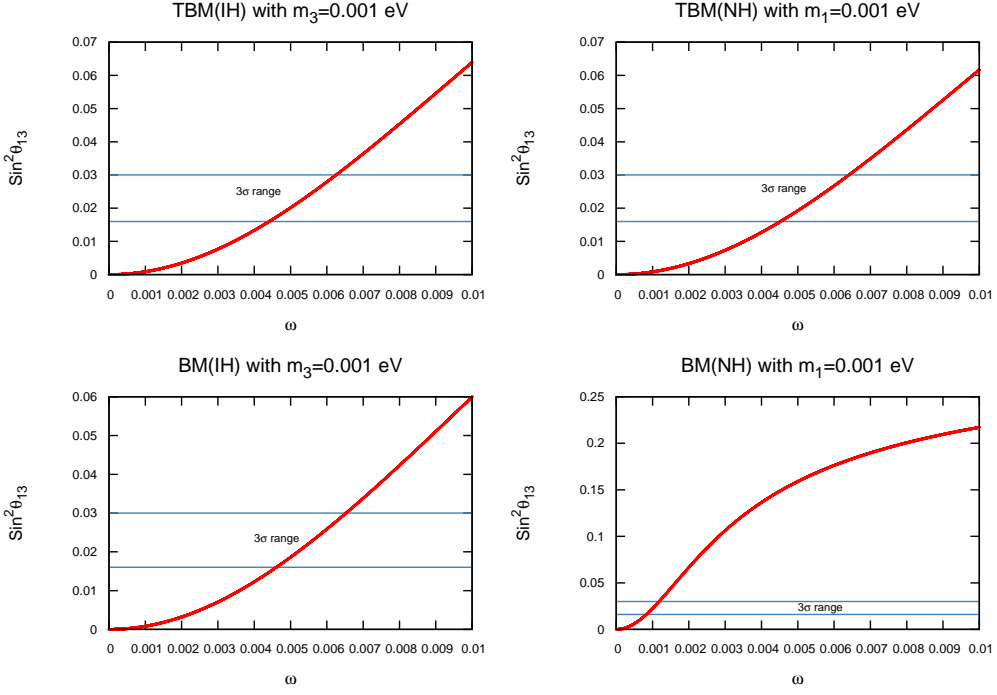}
\caption{Variation of $\sin^2\theta_{13}$ with type II seesaw strength $w$ for BM and TBM 
         with $m_1 (m_3) = 0.07 (0.065)$ eV.}
\label{fig:fig1}
\end{figure}
%
\section{$\mu-\tau$ Symmetric Neutrino Mass Matrix}
\label{sec:mutau}
$\mu-\tau$ symmetric neutrino mass matrix is one of the most widely studied neutrino mixing scenario in the literature. 
In this work, we consider four different types of $\mu-\tau$ symmetric neutrino mass matrix: Bimaximal mixing (BM), 
Tri-bimaximal mixing (TBM), Hexagonal mixing (HM) and Golden Ratio mixing (GRM). These scenarios predict $\theta_{13} = 0, 
\theta_{23} = \frac{\pi}{4}$ whereas the value of $\theta_{12}$ depends upon the particular model. Since $\theta_{13} = 0$ 
has been ruled out by latest neutrino oscillation experiments, the $\mu-\tau$ symmetry has to be broken appropriately 
in order to account for the correct neutrino oscillation data. We assume these four different $\mu-\tau$ symmetric neutrino 
mass matrices to originate from type I seesaw mechanism whereas type II seesaw term acts as a perturbation which breaks 
$\mu-\tau$ symmetry in order to produce the correct neutrino oscillation parameters.

\begin{table}[h!]
\centering
\begin{tabular}{|l|l|l|l|l|}
    \hline
    Parameters (BM) & IH &  NH & IH & NH\\
    \hline \hline
    A & 0.023946 & 0.015114 & 0.0731646&0.0741\\ 
    B & 0.024946 & 0.0141142 & 0.00816462&0.00409996\\ 
    F & 0.00027118& 0.0145033 &0.000163019 &0.00542086\\ 
    \hline
    $m_3$ & 0.001 & 0.0497393 &0.065 &0.0858662\\ 
    $m_2$& 0.0492747 & 0.0087178 &0.0815598 &0.0705337\\ 
    $m_1$ & 0.0485077 & 0.001 & 0.0810987&0.07\\ 
    \hline
    $\sum_i m_i$& 0.0987824 & 0.0594571 &0.22766 &0.22639\\
    \hline
\end{tabular}
\caption{Parametrization of the neutrino mass matrix for BM}
\label{table:BM}
\end{table}

The $\mu-\tau$ symmetric BM type neutrino mass matrix originating from type I seesaw can be parametrised as
\begin{equation}
 m_{LL}=\left(\begin{array}{ccc}
 A+B&F&F\\
 F&A&B\\
 F&B&A
 \end{array}\right)
\end{equation}
This has eigenvalues $m_1=A+B+\sqrt{2}F, m_2=A+B-\sqrt{2}F, m_3=A-B$. It predicts the mixing angles as $\theta_{23}=\theta_{12}=45^o$ 
and $\theta_{13}=0$. It clearly shows that only the first mixing angle $\theta_{23}$ is still allowed from oscillation data whereas 
$\theta_{12} = 45^o$ and $\theta_{13}=0$ have been ruled out experimentally.
\begin{figure}[h]
\centering
\includegraphics[width=0.85\textwidth]{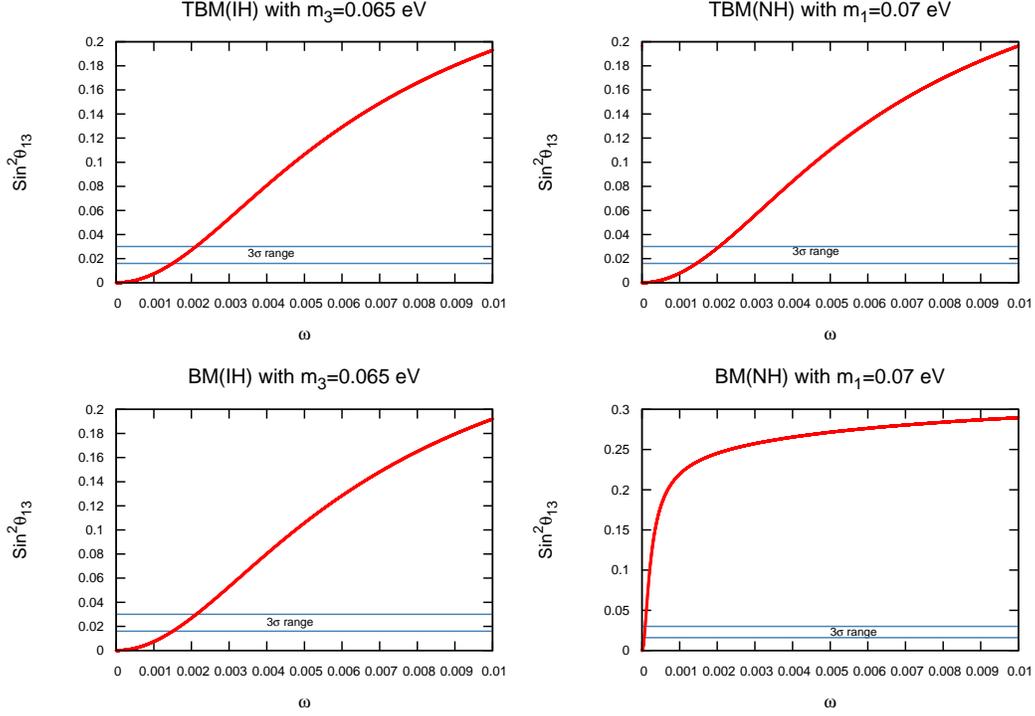}
\caption{$\sin^2\theta_{13}$ with type II seesaw strength $w$ for BM and TBM with $m_1 (m_3) = 0.07 (0.065)$ eV.}
\label{fig:fig2}
\end{figure}

\begin{table}[h!]
\centering
\begin{tabular}{|l|l|l|l|l|}
    \hline
    Parameters (TBM) & IH &  NH & IH & NH\\ 
    \hline \hline 
    A &0.0487942 & 0.0035726 &0.0812524 &0.07017789\\ 
    B & 0.0002555 & 0.0025726 &0.000153696 & 0.00017789\\ 
    F & -0.023769& 0.0243546 &  -0.00804935& 0.007798948\\ 
    \hline
    $m_3$ & 0.001 & 0.0497092 &0.065 &0.0855979\\ 
    $m_2$& 0.0493052 & 0.0087178 &0.0815598 &0.0705337\\ 
    $m_1$ & 0.0485387 & 0.001 & 0.0810987&0.07\\ 
    \hline
    $\sum_i m_i$& 0.098844&0.059427  &0.227657 &0.226132\\
    \hline
\end{tabular}
\caption{Parametrization of the neutrino mass matrix for TBM}
\label{table:TBM}
\end{table}
The $\mu-\tau$ symmetric TBM type neutrino mass matrix originating from type I seesaw can be parametrized as
\begin{equation}
m_{LL}=\left(\begin{array}{ccc}
A& B&B\\
B& A+F & B-F \\
B & B-F & A+F
\end{array}\right)
\label{matrix1}
\end{equation}
which is clearly $\mu-\tau$ symmetric with eigenvalues $m_1 = A-B, \; m_2 = A+2B, \; m_3 = A-B+2F$. It predicts the mixing 
angles as $\theta_{12} \simeq 35.3^o, \; \theta_{23} = 45^o$ and $\theta_{13} = 0$. Although the prediction for first two 
mixing angles are still allowed from oscillation data, $\theta_{13}=0$ has been ruled out experimentally at more than 
$9\sigma$ confidence level.

\begin{table}[h]
\centering
\begin{tabular}{|l|l|l|l|l|}
    \hline
    Parameters (HM) & IH &  NH &IH &NH\\ \hline
    A & 0.048699 & 0.00292945 &0.081214 &0.0701334\\ 
    B & 0.00023485 & 0.00236308 &0.000141179 &0.000163405\\
    F & 0.001 & 0.0497393 &0.065 &0.0858662\\ \hline
    $m_3$ & 0.001 &0.0497393 &0.065 &0.0858662\\ 
    $m_2$& 0.0492747 & 0.0087178 &0.0815598 &0.0705337\\ 
    $m_1$ & 0.0485077 & 0.001 & 0.0810987&0.07\\ \hline
    $\sum_i m_i$& 0.0987824 & 0.0594571 & 0.227658&0.2261957\\
    \hline
\end{tabular}
\caption{Parametrization of the neutrino mass matrix for HM}
\label{table:HM}
\end{table}
\begin{figure}[htb!]
\centering
\includegraphics[width=0.85\textwidth]{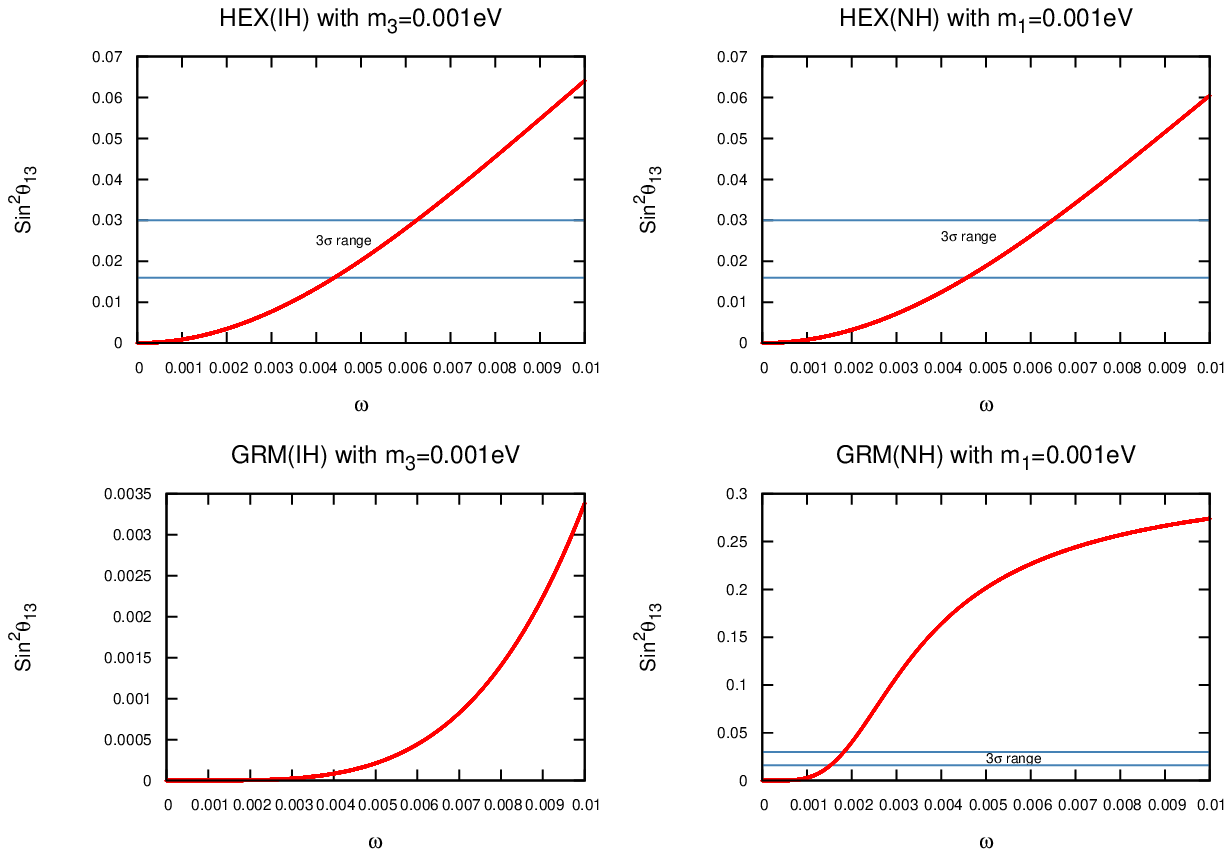}
\caption{$\sin^2\theta_{13}$ with type II seesaw strength $w$ for HM and GRM with $m_1 (m_3) = 0.001$ eV.}
\label{fig:fig3}
\end{figure}
In the same way, the $\mu-\tau$ symmetric Hexagonal mixing (HM) type neutrino mass matrix can be written as
\begin{equation}
 m_{LL}=\left(\begin{array}{ccc}
 A&B&B\\
 B&\frac{1}{2}(A+2\sqrt{\frac{2}{3}}B+F)&\frac{1}{2}(A+2\sqrt{\frac{2}{3}}B-F)\\
 B&\frac{1}{2}(A+2\sqrt{\frac{2}{3}}B-F)&\frac{1}{2}(A+2\sqrt{\frac{2}{3}}B+F)
 \end{array}\right)
 \end{equation}
This has eigenvalues $m_1=\frac{1}{3}(3A-\sqrt{6}B), m_2=A+\sqrt{6}B$ and $m_3=F$ . This predicts the mixing angles 
 to be $\theta_{23}=45^o, \theta_{12}=30^o$ and $\theta_{13}=0$. Oscilation data still allow $\theta_{23}=45^o$ and 
 $\theta_{12}=30^o$ whereas $\theta_{13}=0$ is ruled out.
\begin{table}[h!]
\centering
\begin{tabular}{ | l | l |l|l|l|}
    \hline \hline
    Parameters (GRM) & IH &  NH&IH&NH \\ \hline
    A & 0.0487197& 0.00305254 & 0.0812261&0.0701377\\ 
    B & 0.0002425 & 0.00234835 & 0.000145813 &0.000157545\\ 
    F & 0.0250314 & 0.0270146 &0.0732162 &0.0775182\\ \hline 
    $m_3$ & 0.001 & 0.047655 &0.065 &0.0846759\\ 
    $m_2$& 0.0492747 & 0.0084262 &0.0815598 &0.0704982\\ 
    $m_1$ & 0.0485077 & 0.001 & 0.0810987&0.07\\ \hline  
    $\sum_i m_i$& 0.0987824 & 0.057081 & 0.227658&0.225174\\
    \hline
\end{tabular}
\caption{Parametrization of the neutrino mass matrix for GRM}
\label{table:GRM}
\end{table}
For GRM pattern, the $\mu-\tau$ symmetric neutrino mass matrix can be written as   
 \begin{equation}
m_{LL}=\left(\begin{array}{ccc}
 A&B&B\\
 B&F&A+\sqrt{2}B-F\\
 B&A+\sqrt{2}B-F&F
 \end{array}\right),
 \end{equation}
giving the eigenvalues equal to $m_1=\frac{1}{2}(2A+\sqrt{2}B-\sqrt{10}B), m_2=\frac{1}{2}(2A+\sqrt{2}B+\sqrt{10}B),$ 
and $m_3=-A-\sqrt{2}B+2F$.  This gives rise to neutrino mixing angles as $\theta_{23}=45^o$, $\theta_{12}=31.71^o$ and 
$\theta_{13}=0$. Apart from $\theta_{13}$, the other two mixing angles are still within the 3$\sigma$ range of neutrino 
mixing angles.

\begin{figure}[htb]
\centering
\includegraphics[width=0.85\textwidth]{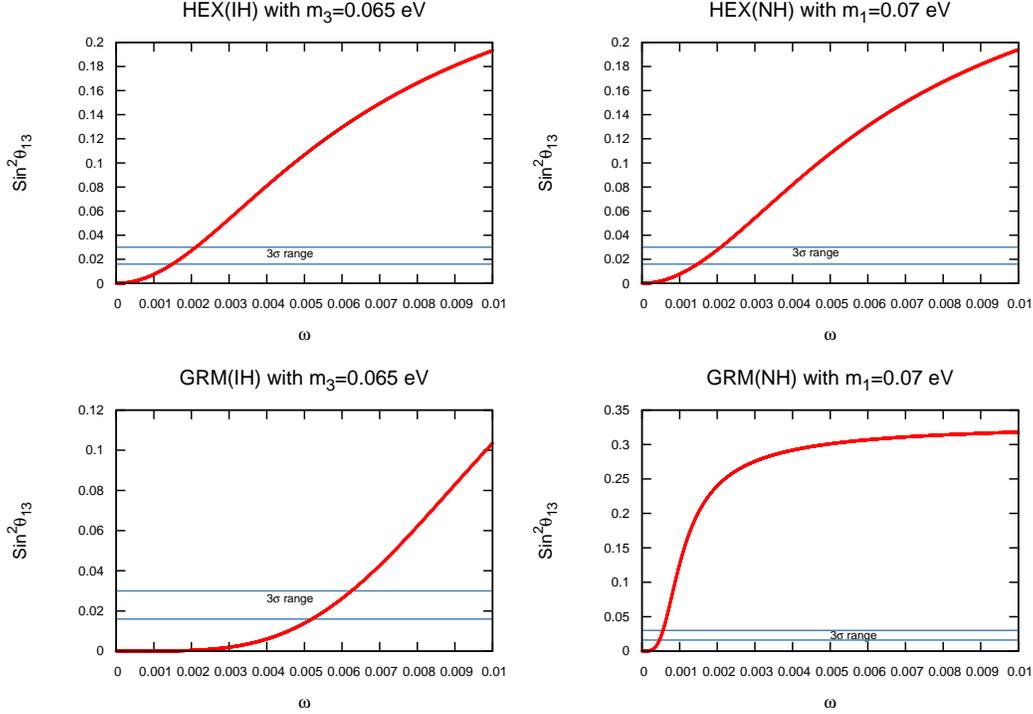}
\caption{$\sin^2\theta_{13}$ with type II seesaw strength $w$ for HM and GRM with $m_1 (m_3) = 0.07 (0.065)$ eV.}
\label{fig:fig4}
\end{figure}

\section{Deviations from $\mu-\tau$ Symmetry}
\label{sec:devmutau}

For simplicity, we assume the type II seesaw mass matrix to be of minimal form while ensuring, at the same time, that it breaks 
the $\mu-\tau$ symmetry in order to generate non-zero $\theta_{13}$. The form of the neutrino mass matrix arising from type II 
seesaw only is taken as
\begin{equation}
m^{II}_{LL}=\left(\begin{array}{ccc}
 0&-w&w\\
 -w&w&0\\
 w&0&-w
 \end{array}\right)
 \label{matrixt2}
 \end{equation}
The structure of this mass matrix although looks ad-hoc, can however, be explained within generic flavor symmetry models like $A_4$. 
Within the framework of seesaw mechanism, neutrino mass and mixing have been extensively studied by many authors using discrete flavor 
symmetries \cite{ish} available in the literature. Among the different discrete flavor symmetry groups, the group of even permutations 
on four elements $A_4$ can naturally explain the $\mu-\tau$ symmetric mass matrix obtained from type I seesaw mechanism. Without going 
into the details of generating a $\mu-\tau$ symmetric mass matrix within $A_4$ models, an exercise performed already by several authors, 
here we briefly outline one possible way of generating the type II seesaw mass matrix (\ref{matrixt2}) within an $A_4$ model. This group 
has $12$ elements having $4$ irreducible representations, with dimensions $n_i$, such that $\sum_i n_i^2=12$. The characters of $4$ 
representations are shown in table \ref{table:character}. The complex number $\omega$ is the cube root of unity. The group $A_4$ has 
four irreducible representations namely, $\bf{1}, \bf{1'}, \bf{1''}$ and $\bf{3}$. In generic $A_4$ models, the $SU(2)_L$ lepton doublets 
$l = (l_e, l_{\mu}, l_{\tau})$ are assumed to transform as triplet $\bf{3}$ under $A_4$ whereas the $SU(2)_L$ singlet charged leptons 
$e^c, \mu^c, \tau^c$ transform as $\bf{1}, \bf{1'}, \bf{1''}$ respectively. In type I seesaw scenarios, the $SU(2)_L$ singlet right handed 
neutrinos $\nu^c$ transform as a triplet under $A_4$. Since we are trying to explain the structure of type II term only, we confine our 
discussion to the lepton doublets only. We introduce three scalars $\zeta_1, \zeta_2, \zeta_3$ transforming as $\bf{1}, \bf{1'}, \bf{1''}$ 
under $A_4$. The $SU(2)_L$ triplet Higgs field $\Delta_L$ is assumed to be a singlet under $A_4$. Thus the type II seesaw term 
can be written as 
$$ \mathcal{L}^{II} = f l l (\zeta_1+\zeta_2+\zeta_3) \Delta_L/\Lambda$$
where $\Lambda$ is the cutoff scale and $f$ is a dimensionless coupling constant.

\begin{table}[ht]
\centering
\caption{Character table of $A_4$}
\vspace{0.5cm}
\begin{tabular}{ccccc}
 \hline
   Class & $\chi^{(1)}$ &  $\chi^{(2)}$ &  $\chi^{(3)}$&  $\chi^{(4)}$\\ \hline \hline
   $C_1$ & 1&1&1&3 \\ \hline 
    $C_2$& 1&$\omega$&$\omega^2$&0 \\  
$C_3$ &1&$\omega^2$&$\omega$&0\\ 
$C_4$ &1&1&1&-1\\  \hline
\end{tabular}
\label{table:character}
\end{table}
The decomposition of the $ll\zeta_{1,2,3}$ terms into $A_4$ singlet gives
$$ ll\zeta_1 = (l_el_e+l_{\mu}l_{\tau}+l_{\tau}l_{\mu}) \zeta_1 $$
$$ ll\zeta_2 = (l_{\mu}l_{\mu}+l_{e}l_{\tau}+l_{\tau}l_{e}) \zeta_2 $$
$$ ll\zeta_3 = (l_{\tau}l_{\tau}+l_{e}l_{\mu}+l_{\mu}l_{e}) \zeta_3 $$
Assuming the vacuum alignments of the scalars as $ \langle \zeta_1 \rangle = 0, \langle \zeta_2 \rangle = \Lambda, \langle \zeta_{3} 
\rangle = -\Lambda$, we obtain the type II seesaw contribution to neutrino mass as
\begin{equation}
m^{II}_{LL}=\left(\begin{array}{ccc}
0& -f \langle \delta^0_L \rangle & f \langle \delta^0_L \rangle\\
-f \langle \delta^0_L \rangle & f \langle \delta^0_L \rangle & 0 \\
f \langle \delta^0_L \rangle & 0& -f \langle \delta^0_L \rangle
\end{array}\right)
\label{matrixt21}
\end{equation}
which has the same form as (\ref{matrixt2}) if we denote $f \langle \delta^0_L \rangle = fv_L$ as $w$. We adopt this minimal 
structure of the type II seesaw mass matrix for our numerical analysis.

\section{CP violation and Leptogenesis}
\label{sec:lepto}
Leptogenesis is one of the most widely studied mechanisms to generate the observed baryon asymmetry of the Universe by creating an 
asymmetry in the leptonic sector first, which subsequently gets converted into baryon asymmetry through $B+L$ violating sphaleron 
processes during electroweak phase transition. Since quark sector CP violation is not sufficient for producing observed baryon 
asymmetry, a framework explaining non-zero $\theta_{13}$ and leptonic CP phase could not only give a better picture of leptonic 
flavor structure, but also the origin of matter-antimatter asymmetry. 

In a model with both type I and type II seesaw mechanisms at work, there are two possible sources of lepton asymmetry: either the CP 
violating decay of the lightest right handed neutrino or that of scalar triplet. Recently, such a work was performed in \cite{leptodborah} 
where the contributions of type I and type II seesaw to baryon asymmetry were calculated without assuming any specific symmetries in the 
type I or type II seesaw matrices. In another work \cite{dbijmpa}, type II seesaw was considered to be the origin of non-zero $\theta_{13}$ 
and non-trivial Dirac $CP$ phase simultaneously and baryon asymmetry was calculated taking contribution only from the type II seesaw term. 
In the present work, both type I and type II seesaw mass matrices are real and hence the diagonalizing matrix $U_{\nu}$ of neutrino mass 
matrix is also real giving rise to trivial values of Dirac CP phase. Thus, the only remaining source of CP violation in leptonic sector 
is the charged lepton sector. We note that the Pontecorvo-Maki-Nakagawa-Sakata (PMNS) leptonic mixing matrix is related to the diagonalizing 
matrices of neutrino and charged lepton mass matrices $U_{\nu}, U_l$ respectively, as
\begin{equation}
U_{\text{PMNS}} = U^{\dagger}_l U_{\nu}
\label{pmns0}
\end{equation}
The PMNS mixing matrix can be parametrized as
\begin{equation}
U_{\text{PMNS}}=\left(\begin{array}{ccc}
c_{12}c_{13}& s_{12}c_{13}& s_{13}e^{-i\delta}\\
-s_{12}c_{23}-c_{12}s_{23}s_{13}e^{i\delta}& c_{12}c_{23}-s_{12}s_{23}s_{13}e^{i\delta} & s_{23}c_{13} \\
s_{12}s_{23}-c_{12}c_{23}s_{13}e^{i\delta} & -c_{12}s_{23}-s_{12}c_{23}s_{13}e^{i\delta}& c_{23}c_{13}
\end{array}\right) 
\label{matrixPMNS}
\end{equation}
where $c_{ij} = \cos{\theta_{ij}}, \; s_{ij} = \sin{\theta_{ij}}$ and $\delta$ is the Dirac CP phase. Our goal is to generate correct 
values of neutrino mixing angles including non-zero $\theta_{13}$ with the combination of type I and type II seesaw. Since, neutrino 
mass matrix is real without any phase, its diagonalizing matrix $U_{\nu}$ is also real and takes the form of $U_{\text{PMNS}}$ after 
setting $\delta$ to zero. Thus, the charged lepton mass diagonalizing matrix $U_l$, the only source of non-zero CP phase $\delta$ 
can be written as
\begin{equation}
U_l=\left(\begin{array}{ccc}
c^2_{13}+e^{i\delta}s^2_{13}& (1-e^{-i\delta})c_{13}s_{13}s_{23}& (1-e^{-i\delta})c_{13}s_{13}c_{23}\\
(-1+e^{i\delta})c_{13}s_{13}s_{23}& c^2_{13}+s^2_{13}(c^2_{23}+e^{-i\delta}s^2_{23}) & (-1+e^{-i\delta})c_{23}s^2_{13}s_{23} \\
(-1+e^{i\delta})c_{13}s_{13}c_{23} & (-1+e^{-i\delta})c_{23}s^2_{13}s_{23}& c^2_{13}+s^2_{13}(s^2_{23}+e^{-i\delta}c^2_{23})
\end{array}\right) 
\label{matrixUl}
\end{equation}
We derive this form of $U_l$ such that $U^{\dagger}_l U_{\nu}$ gives the desired form of PMNS mixing matrix (\ref{matrixPMNS}). 
If we assume that this matrix $U_l$ also diagonalizes the Dirac neutrino mass matrix $m_{LR}$, the $CP$ phase originating in the 
charged lepton sector can affect the lepton asymmetry as we discuss below.

\begin{figure}[h]
\begin{center}
\includegraphics[width=0.85\textwidth]{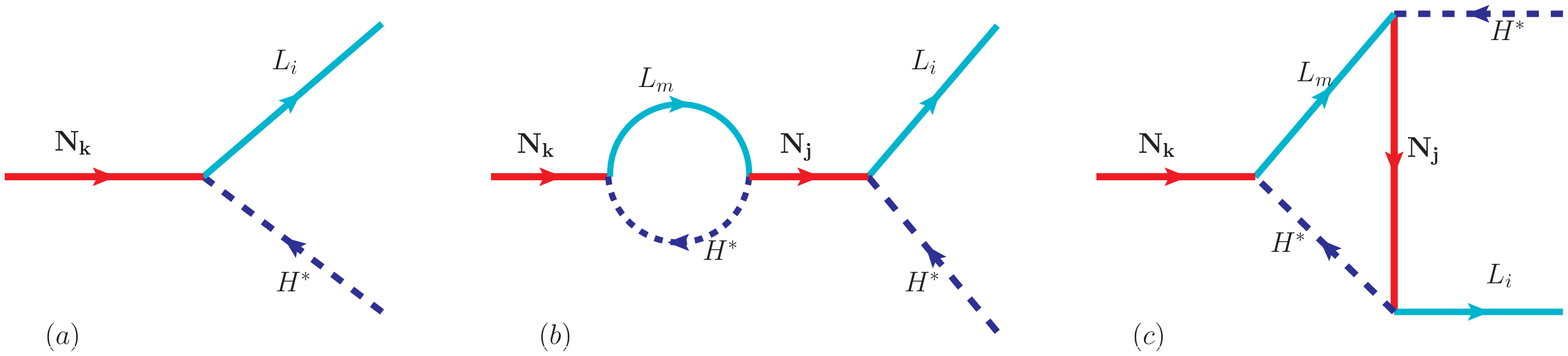}
\end{center}
\caption{Right handed neutrino decay}
\label{fig0001}
\end{figure}
\begin{figure}
\begin{center}
\includegraphics[width=0.5\textwidth]{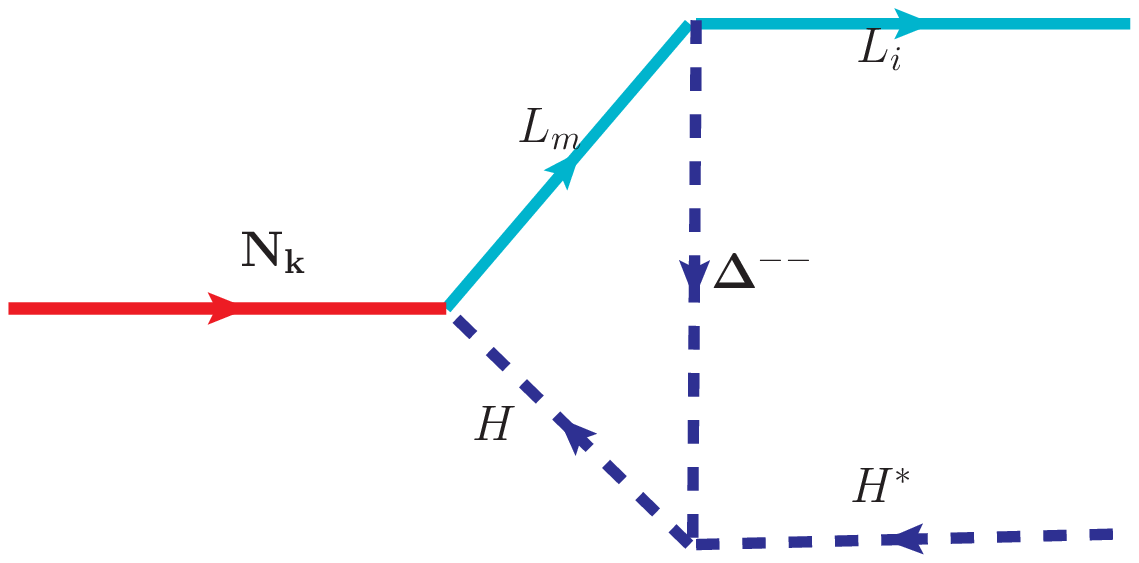}
\end{center}
\caption{Right handed neutrino decay}
\label{fig01}
\end{figure}

In our work we are considering CP-violating out of equilibrium decay of heavy RH neutrinos in to Higgs and lepton 
within the framework of dominant type I and sub-dominant type II seesaw mechanism. In principle, the decay of Higgs triplet 
having masses few hundred GeV can contribute to the CP-asymmetry in the lepton sector having prominent gauge interaction
along with the as usual CP-asymmetry due to heavy ($> 10^{9}$ GeV) right-handed neutrino decays without having any gauge 
interaction. The wash-out factors in case of CP-asymmetry due to Triplet decay is large and thus, the net CP-asymmetry is 
negligible. For simplicity we consider only the right handed neutrino decay as a source of lepton asymmetry and neglect 
the contribution coming from triplet decay. The lepton asymmetry from the decay of right handed neutrino into leptons 
and Higgs scalar is given by
\begin{equation}
\epsilon_{N_k} = \sum_i \frac{\Gamma(N_k \rightarrow L_i +H^*)-\Gamma (N_k \rightarrow \bar{L_i}+H)}{\Gamma(N_k \rightarrow L_i +H^*)
+\Gamma (N_k \rightarrow \bar{L_i}+H)}
\end{equation}
In a hierarchical pattern for right handed neutrinos $M_{2,3} \gg M_1$, it is sufficient to consider the lepton asymmetry produced by 
the decay of lightest right handed neutrino $N_1$ decay. In a type I seesaw framework where the particle content is just the standard 
model with three additional right handed neutrinos, the lepton asymmetry is generated through the decay processes shown in figure 
\ref{fig0001}. In the presence of type II seesaw, $N_1$ can also decay through a virtual triplet as can be seen 
in figure \ref{fig01}. Following the notations of \cite{joshipura}, the lepton asymmetry arising from the decay of $N_1$ in the 
presence of type I seesaw only can be written as
\begin{eqnarray}
\epsilon^{\alpha}_1 &=& \frac{1}{8\pi v^2}\frac{1}{(m^{\dagger}_{LR}m_{LR})_{11}} \sum_{j=2,3} \text{Im}[(m^*_{LR})_{\alpha 1}
(m^{\dagger}_{LR}m_{LR})_{1j}(m_{LR})_{\alpha j}]g(x_j) \nonumber \\&& + \frac{1}{8\pi v^2}\frac{1}{(m^{\dagger}_{LR}m_{LR})_{11}} 
\sum_{j=2,3} \text{Im}[(m^*_{LR})_{\alpha 1}(m^{\dagger}_{LR}m_{LR})_{j1}(m_{LR})_{\alpha j}]\frac{1}{1-x_j}
\label{eps1}
\end{eqnarray}
where $v = 174 \; \text{GeV}$ is the vev of the Higgs bidoublets responsible for breaking the electroweak symmetry, $$ g(x) = \sqrt{x} 
\left ( 1+\frac{1}{1-x}-(1+x)\text{ln}\frac{1+x}{x} \right) $$and $x_j = M^2_j/M^2_1$. The second term in the expression for $\epsilon^{\alpha}_1$ 
above vanishes when summed over all the flavors $\alpha = e, \mu, \tau$. The sum over flavors is given by
\begin{equation}
\epsilon_1 = \frac{1}{8\pi v^2}\frac{1}{(m^{\dagger}_{LR}m_{LR})_{11}}\sum_{j=2,3} \text{Im}[(m^{\dagger}_{LR}m_{LR})^2_{1j}]g(x_j)
\label{noflavor}
\end{equation}
\begin{figure}
\centering
\includegraphics[width=0.85\textwidth]{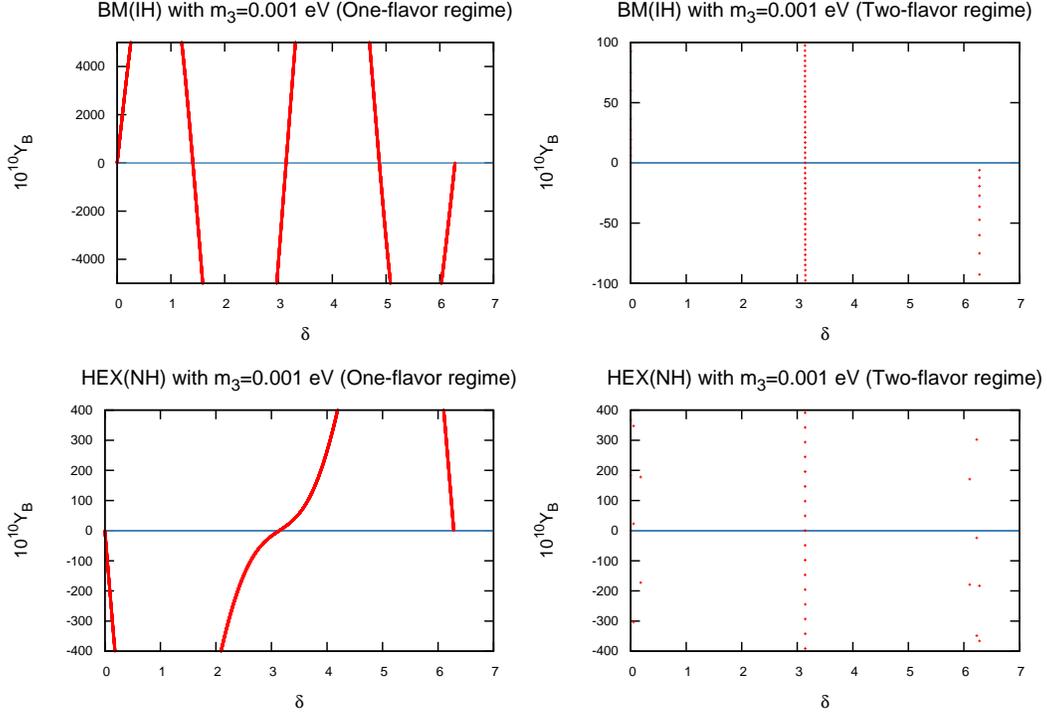}
\caption{Variation of baryon asymmetry with $\delta$ for BM and HM}
\label{fig:fig29}
\end{figure}
\begin{figure}
\centering
\includegraphics[width=0.85\textwidth]{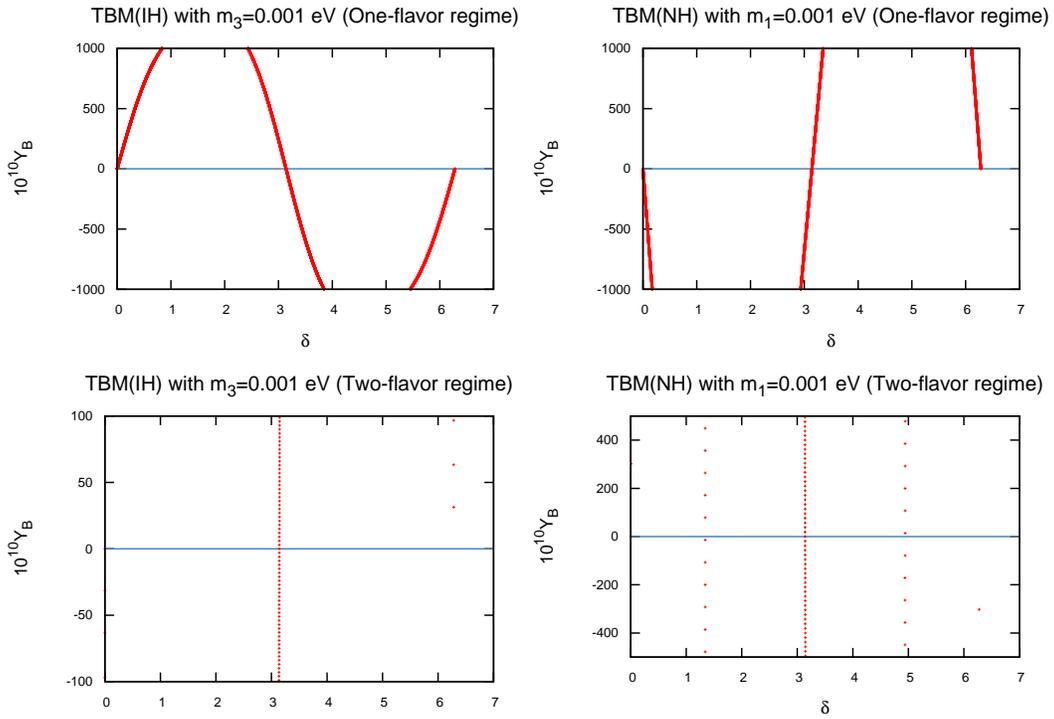}
\caption{Variation of baryon asymmetry with $\delta$ for TBM}
\label{fig:fig30}
\end{figure}
\begin{figure}
\centering
\includegraphics[width=0.85\textwidth]{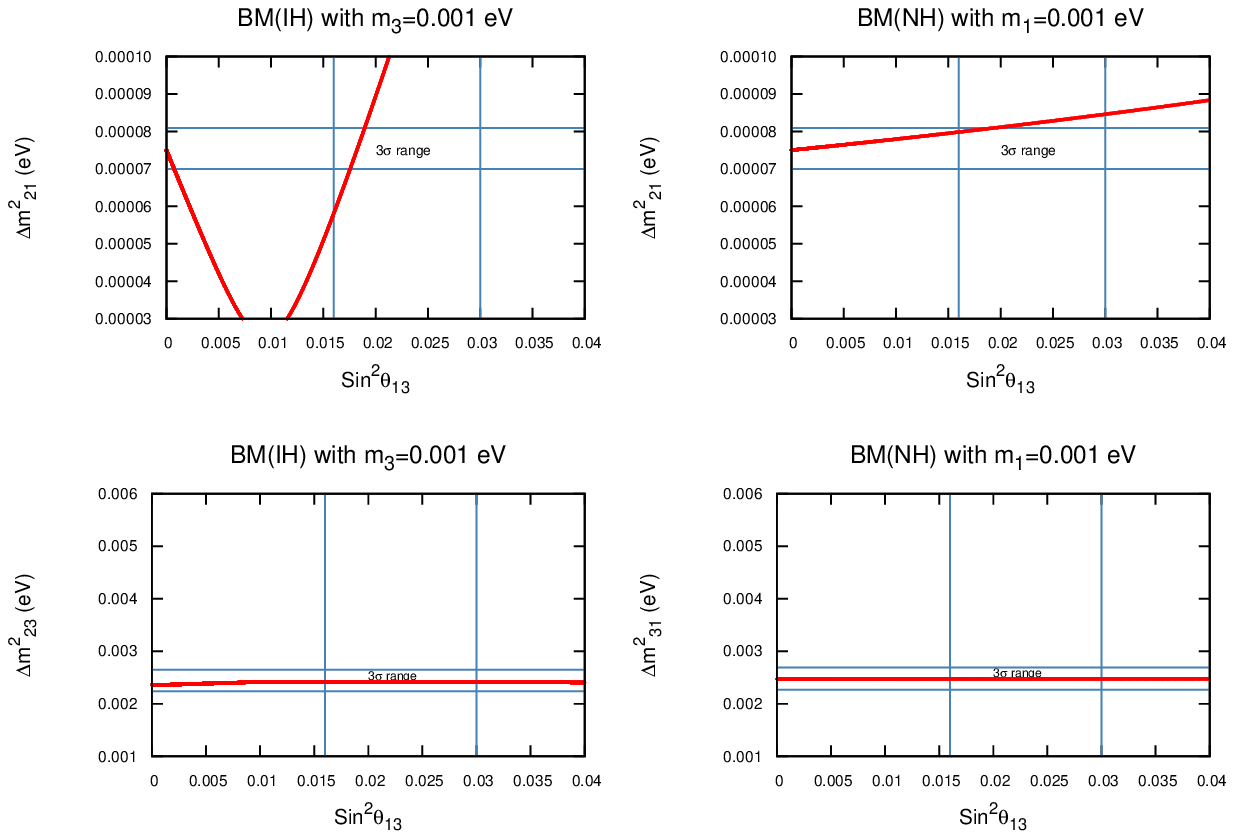}
\caption{$\Delta m^2_{21}$, $\Delta m^2_{23}$, $\Delta m^2_{31}$ with $\sin^2\theta_{13}$ for BM with $m_1 (m_3) = 0.001$ eV.}
\label{fig:fig5}
\end{figure}
\begin{figure}
\centering
\includegraphics[width=0.85\textwidth]{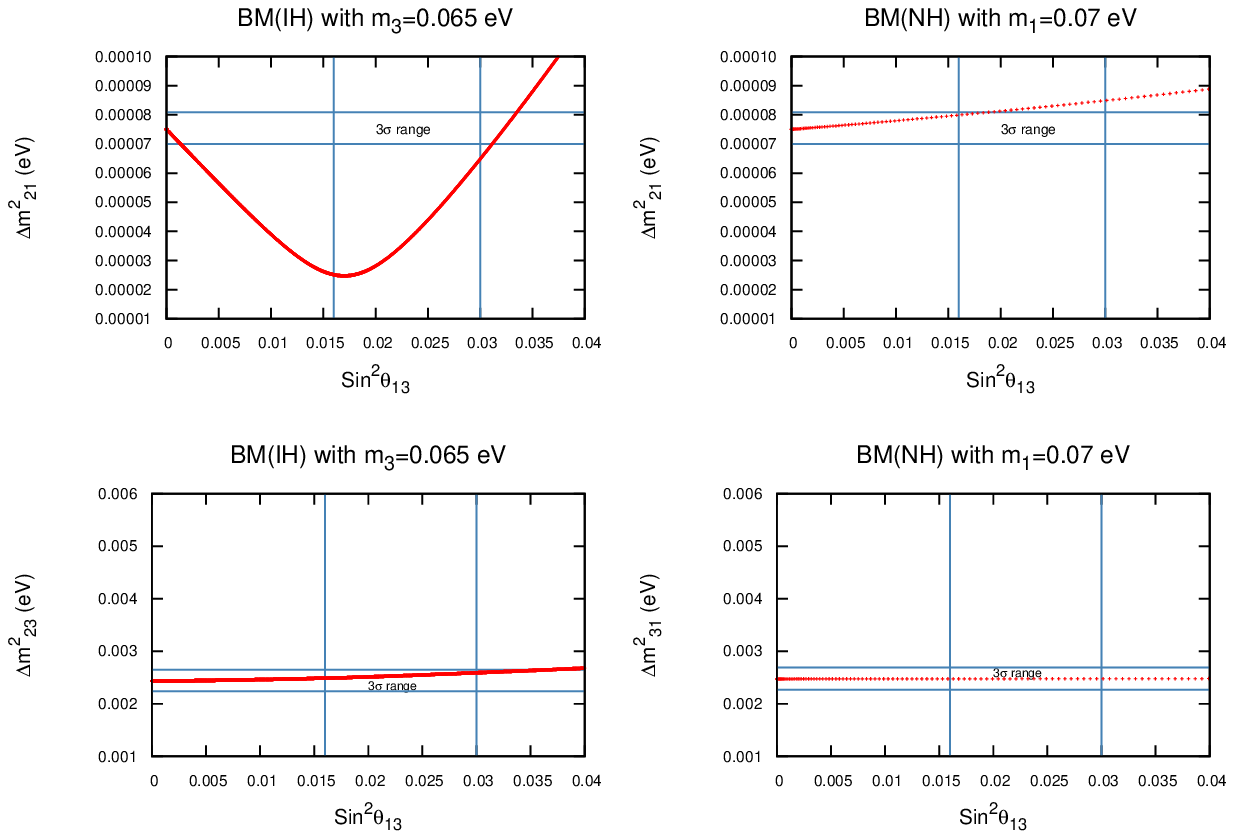}
\caption{$\Delta m^2_{21}$, $\Delta m^2_{23}$, $\Delta m^2_{31}$ with $\sin^2\theta_{13}$ for BM with $m_1 (m_3) = 0.07 (0.065)$ eV. }
\label{fig:fig6}
\end{figure}
\begin{figure}[t]
\centering
\includegraphics[width=0.85\textwidth]{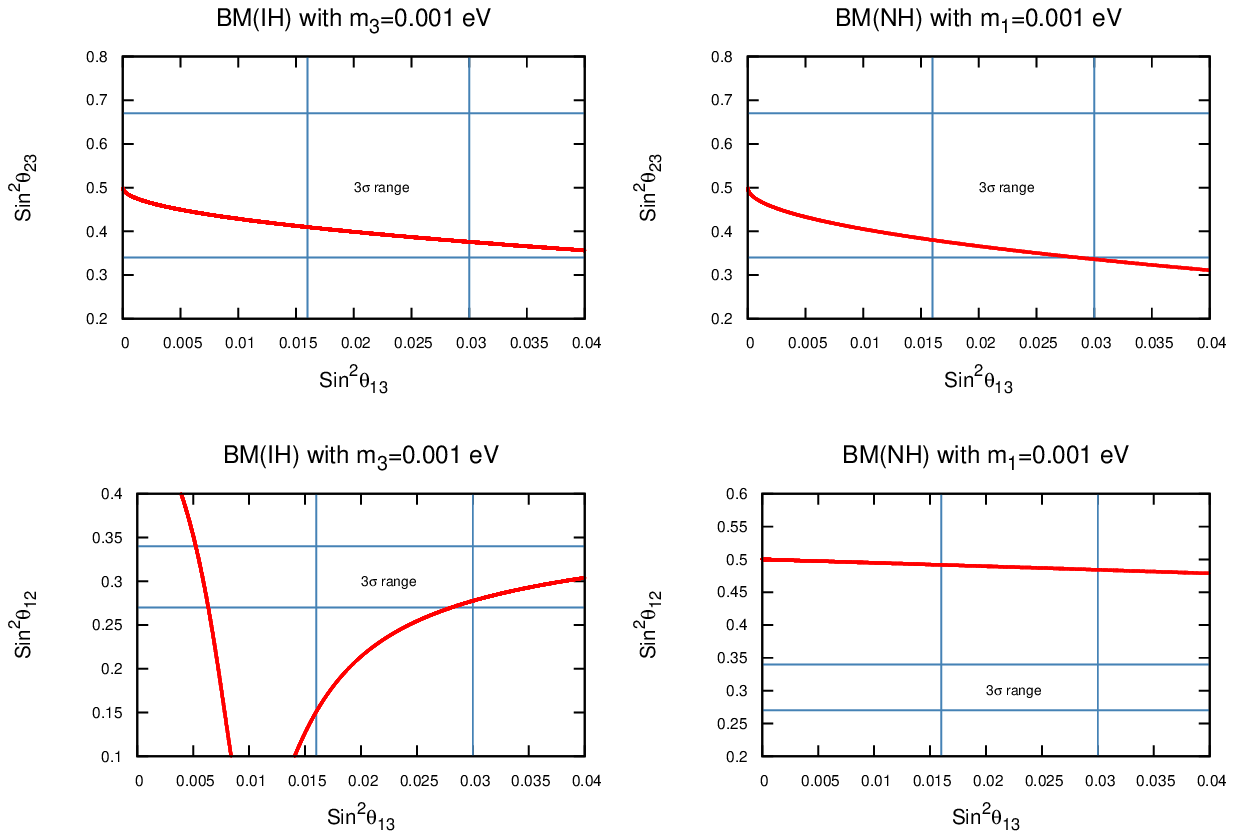}
\caption{$\sin^2\theta_{23}$, $\sin^2\theta_{12}$ with $\sin^2\theta_{13}$ for BM with $m_1 (m_3) = 0.001$ eV.}
\label{fig:fig7}
\end{figure}
\begin{figure}
\centering
\includegraphics[width=0.85\textwidth]{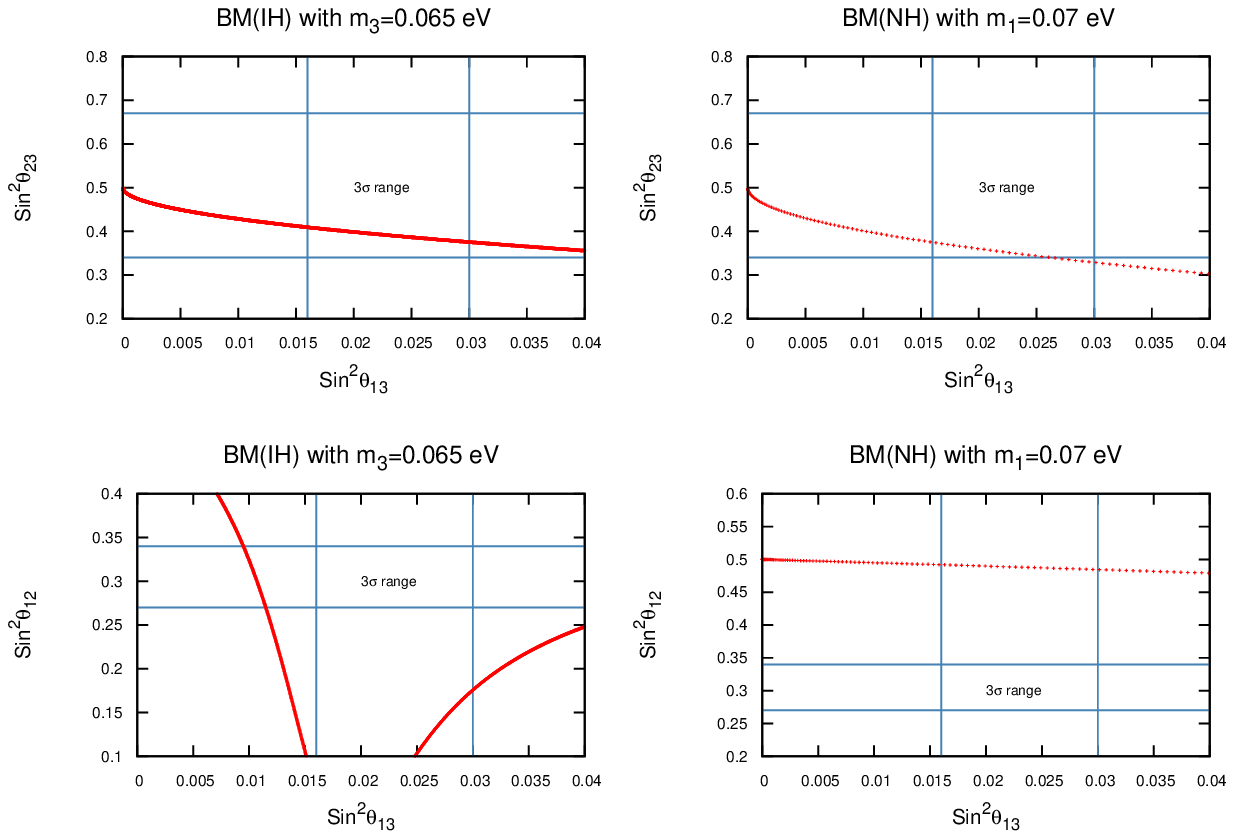}
\caption{$\sin^2\theta_{23}$, $\sin^2\theta_{12}$ with $\sin^2\theta_{13}$ for BM with $m_1 (m_3) = 0.07 (0.065)$ eV.}
\label{fig:fig8}
\end{figure}
\begin{figure}
\centering
\includegraphics[width=0.85\textwidth]{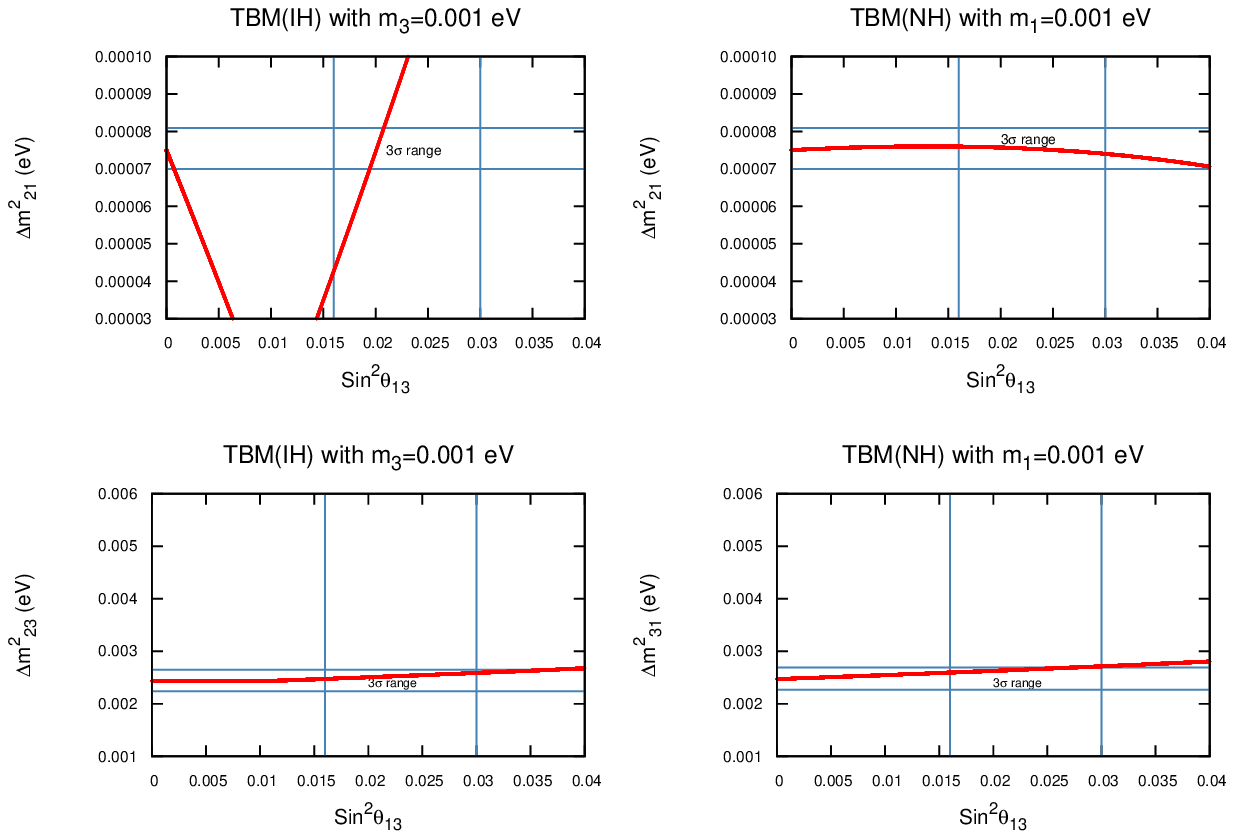}
\caption{$\Delta m^2_{21}$, $\Delta m^2_{23}$, $\Delta m^2_{31}$ with $\sin^2\theta_{13}$ for TBM with $m_1 (m_3) = 0.001$ eV.}
\label{fig:fig9}
\end{figure}
\begin{figure}
\centering
\includegraphics[width=0.85\textwidth]{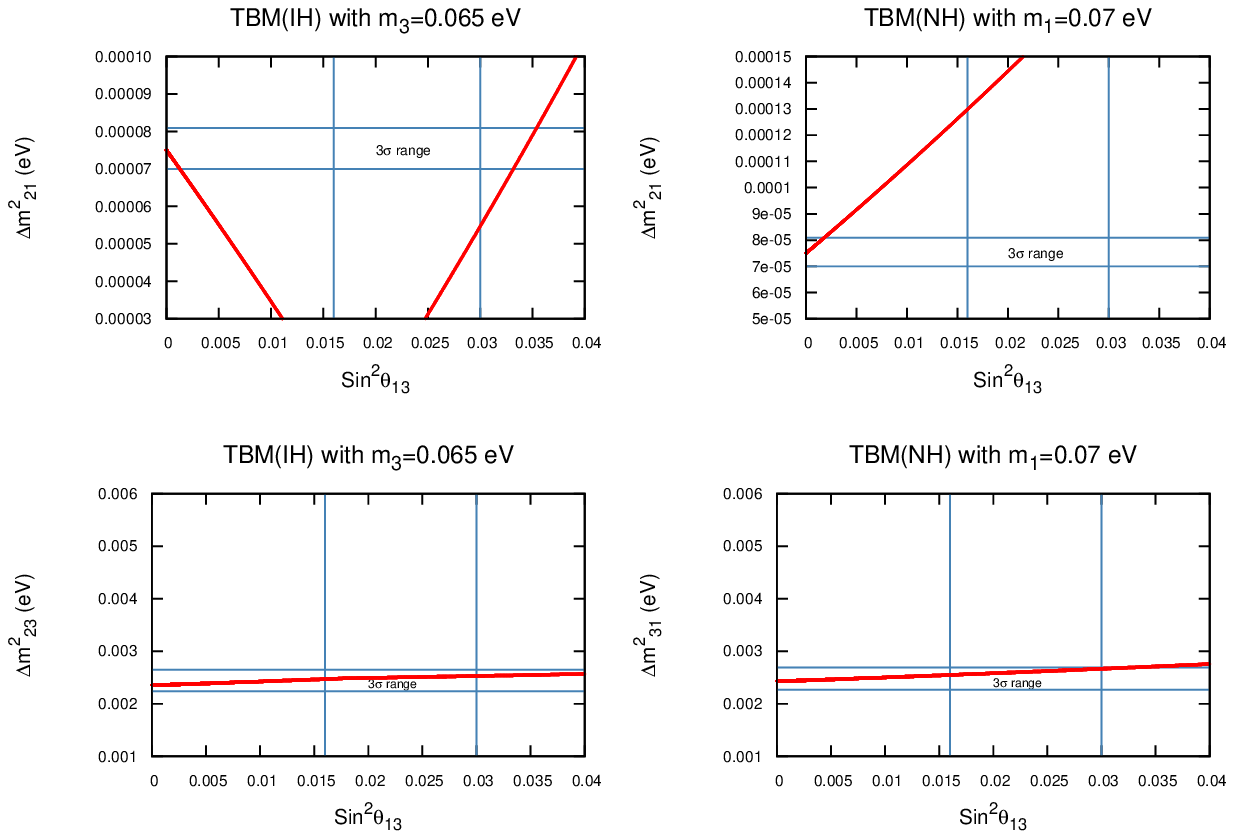}
\caption{$\Delta m^2_{21}$, $\Delta m^2_{23}$, $\Delta m^2_{31}$ with $\sin^2\theta_{13}$ for TBM with $m_1 (m_3) = 0.07 (0.065)$ eV.}
\label{fig:fig10}
\end{figure}
\begin{figure}
\centering
\includegraphics[width=0.85\textwidth]{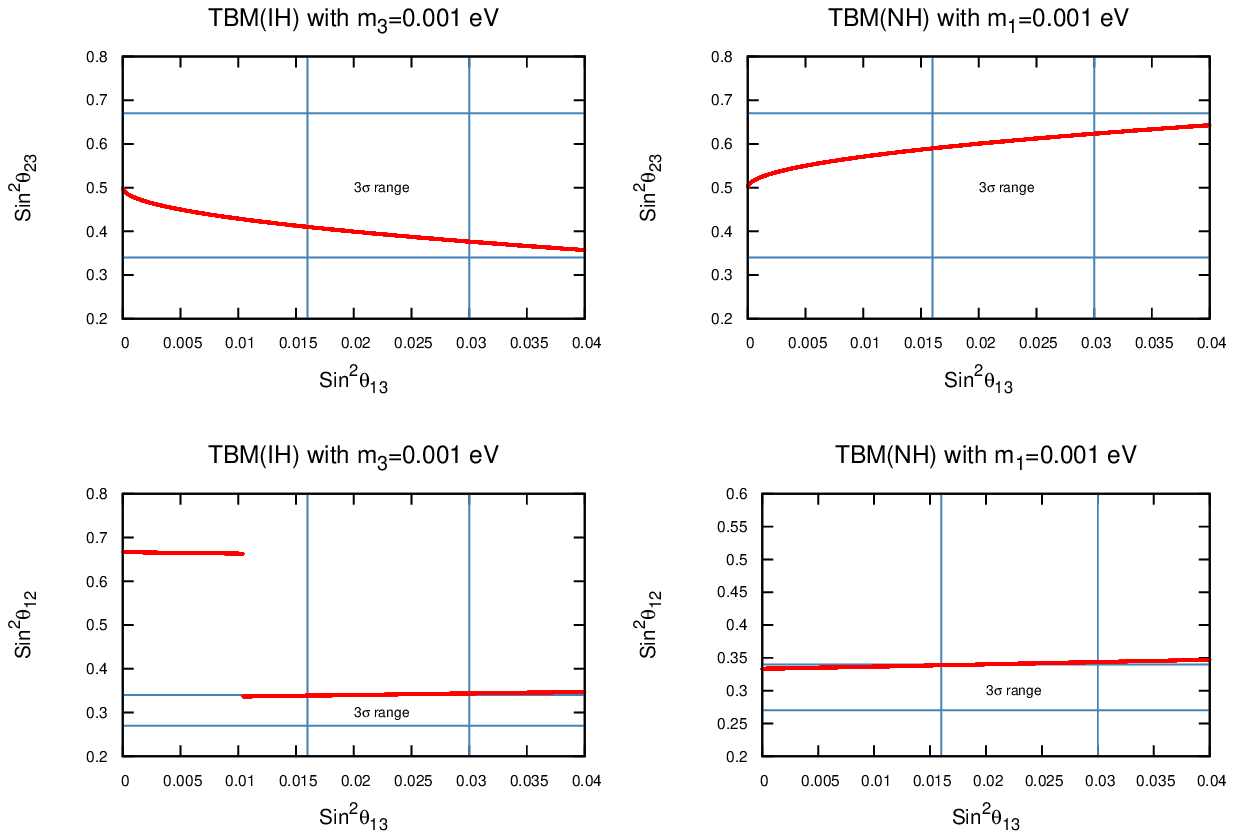}
\caption{$\sin^2\theta_{23}$, $\sin^2\theta_{12}$ with $\sin^2\theta_{13}$ for TBM with $m_1 (m_3) = 0.001$ eV.}
\label{fig:fig11}
\end{figure}
\begin{figure}
\centering
\includegraphics[width=0.85\textwidth]{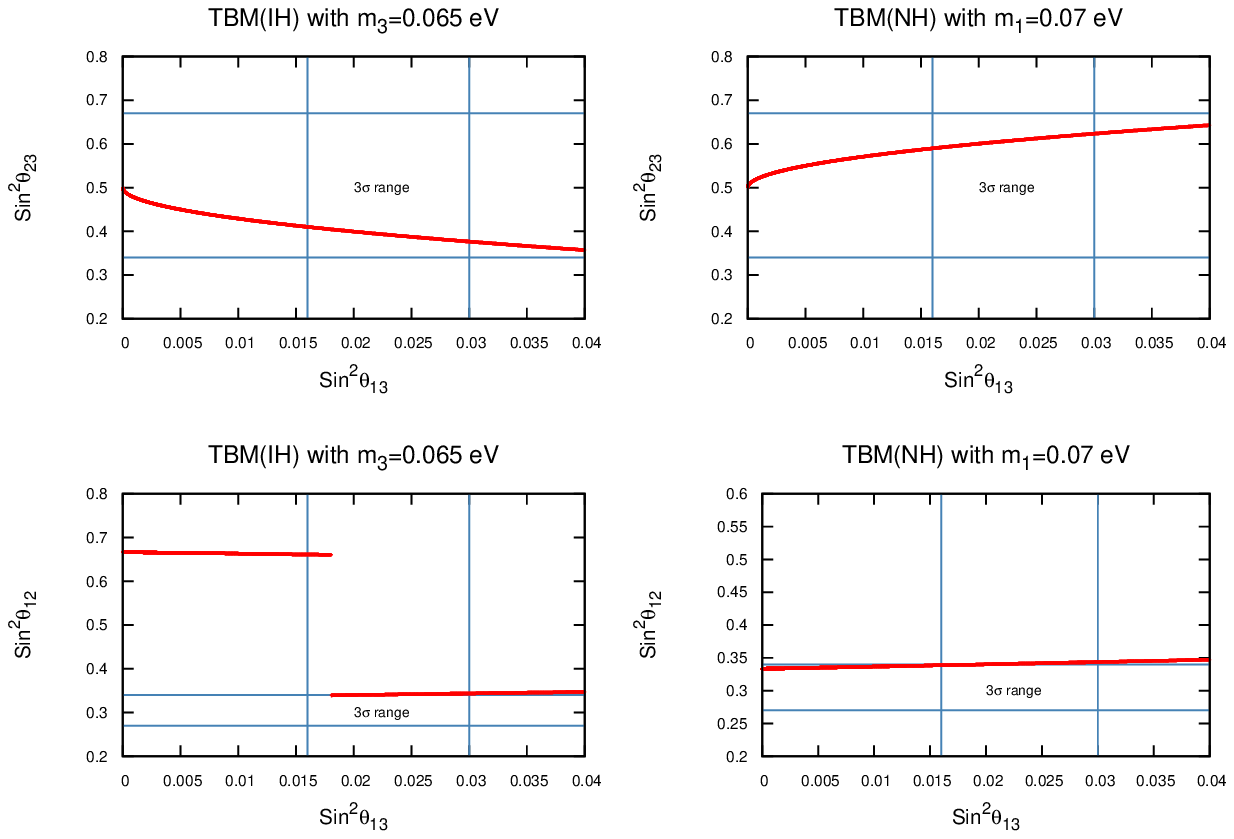}
\caption{$\sin^2\theta_{23}$, $\sin^2\theta_{12}$ with $\sin^2\theta_{13}$ for TBM with $m_1 (m_3) = 0.07 (0.065)$ eV.}
\label{fig:fig12}
\end{figure}
\begin{figure}
\centering
\includegraphics[width=0.85\textwidth]{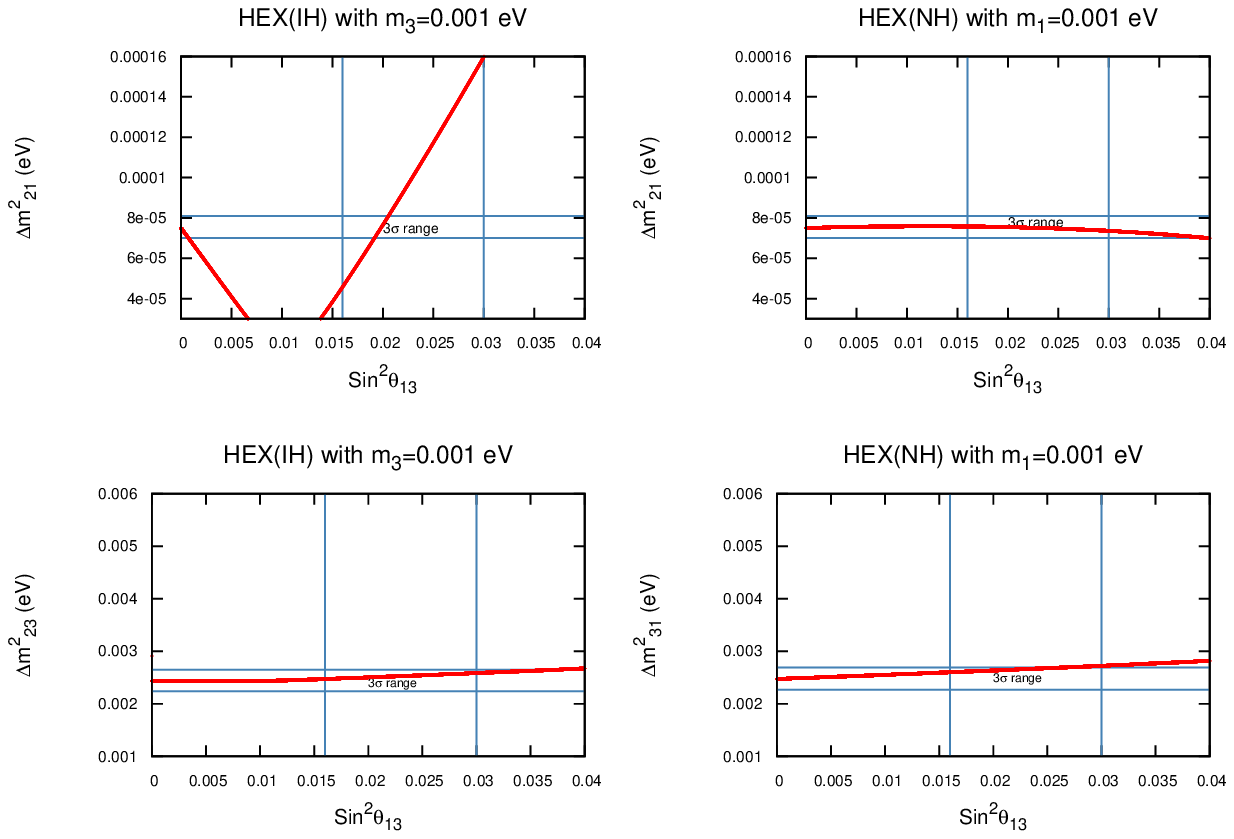}
\caption{$\Delta m^2_{21}$, $\Delta m^2_{23}$, $\Delta m^2_{31}$ with $\sin^2\theta_{13}$ for HM with $m_1 (m_3) = 0.001$ eV.}
\label{fig:fig13}
\end{figure}

\begin{figure}
\centering
\includegraphics[width=0.85\textwidth]{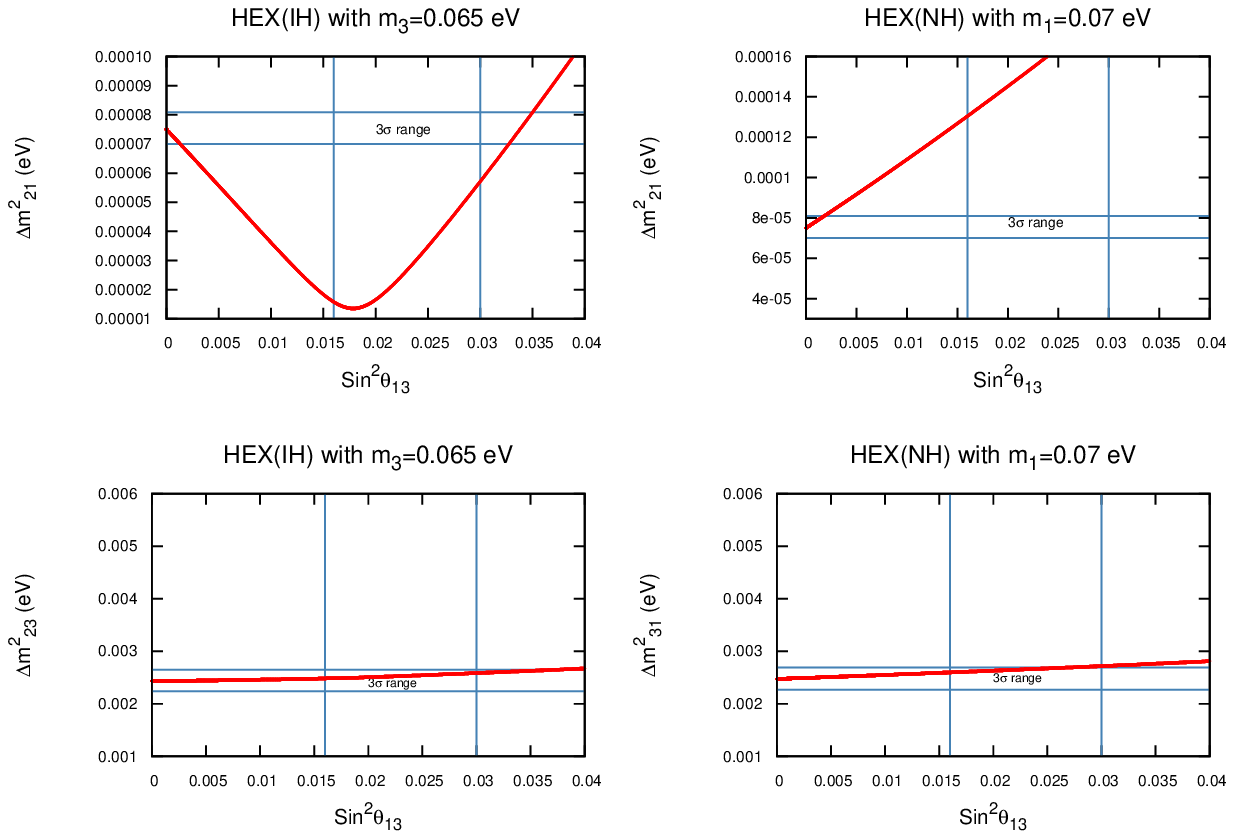}
\caption{$\Delta m^2_{21}$, $\Delta m^2_{23}$, $\Delta m^2_{31}$ with $\sin^2\theta_{13}$ for HM with $m_1 (m_3) = 0.07 (0.065)$ eV.}
\label{fig:fig14}
\end{figure}

\begin{figure}
\centering
\includegraphics[width=0.85\textwidth]{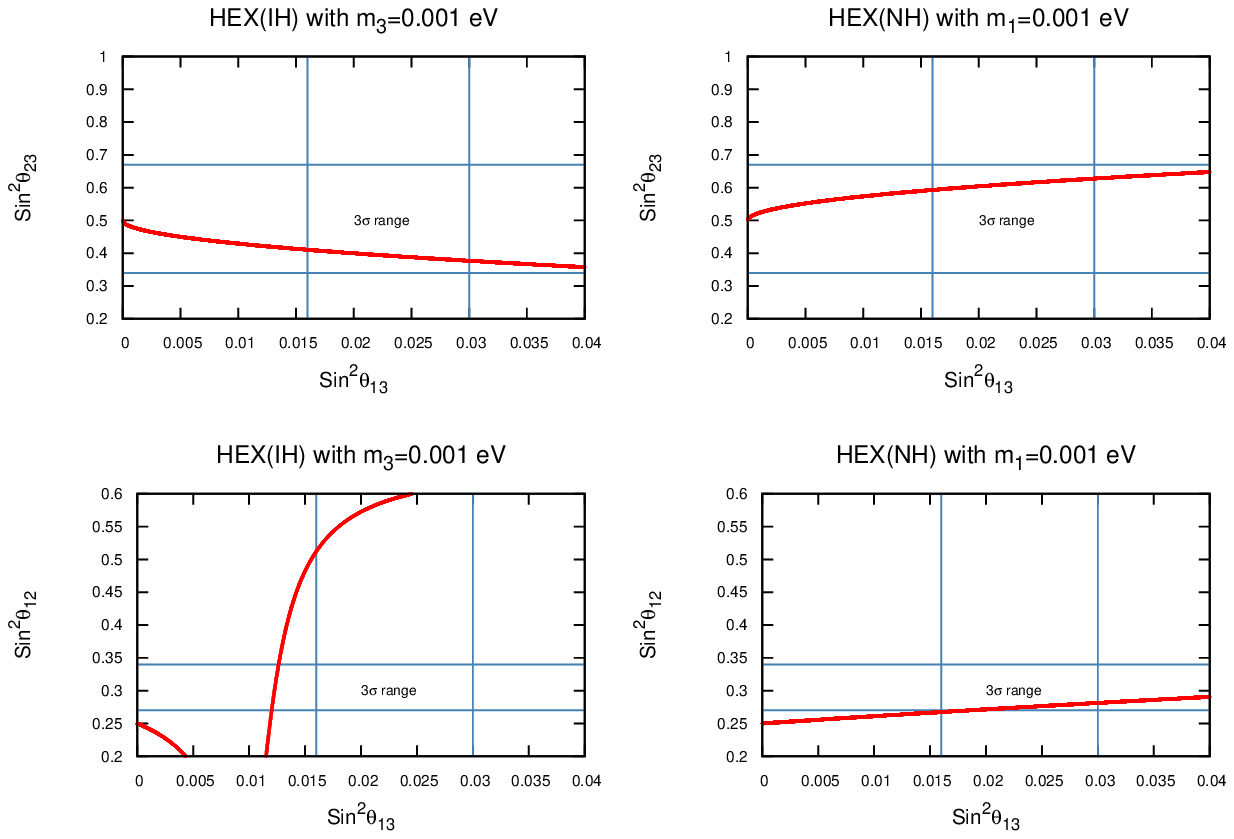}
\caption{$\sin^2\theta_{23}$, $\sin^2\theta_{12}$ with $\sin^2\theta_{13}$ for HM with $m_1 (m_3) = 0.001$ eV.}
\label{fig:fig15}
\end{figure}
 
\begin{figure}
\centering
\includegraphics[width=0.85\textwidth]{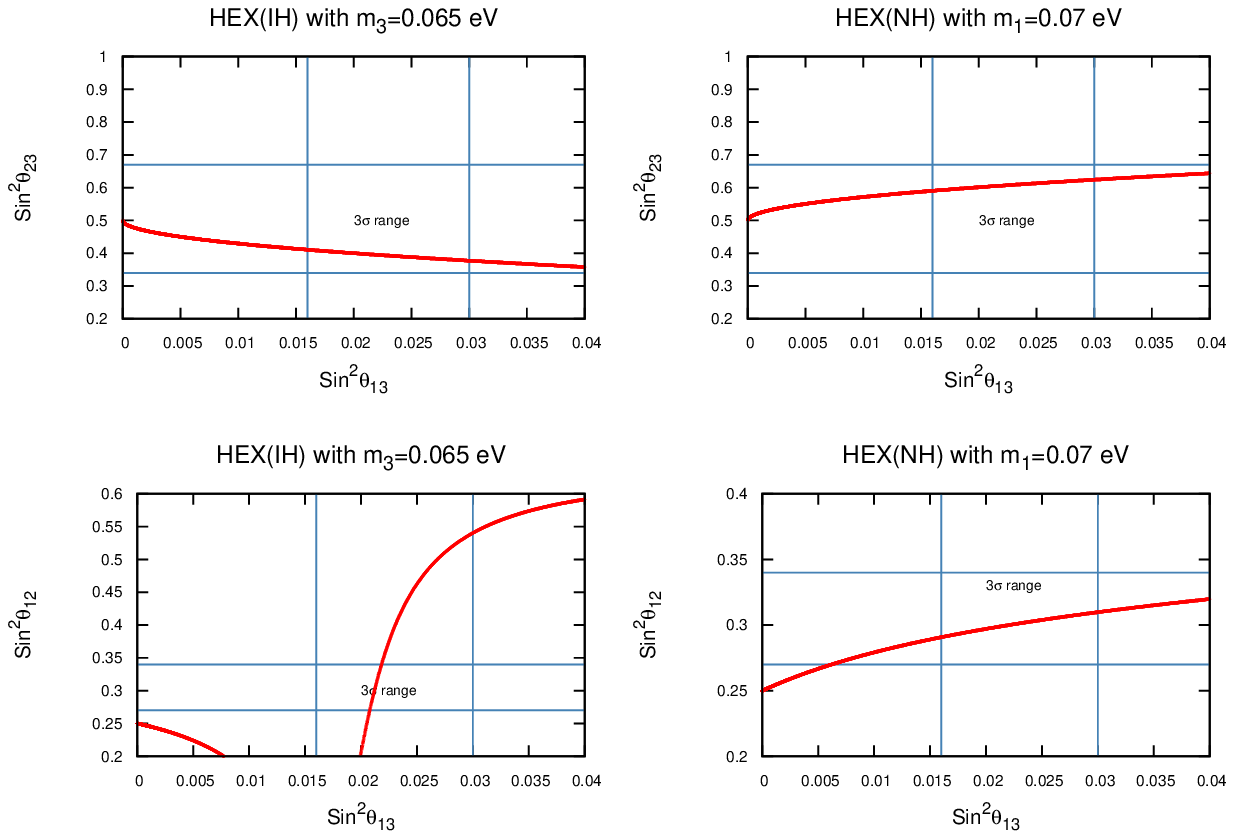}
\caption{$\sin^2\theta_{23}$, $\sin^2\theta_{12}$ with $\sin^2\theta_{13}$ for HM with $m_1 (m_3) = 0.07 (0.065)$ eV.}
\label{fig:fig16}
\end{figure}

\begin{figure}
\centering
\includegraphics[width=0.85\textwidth]{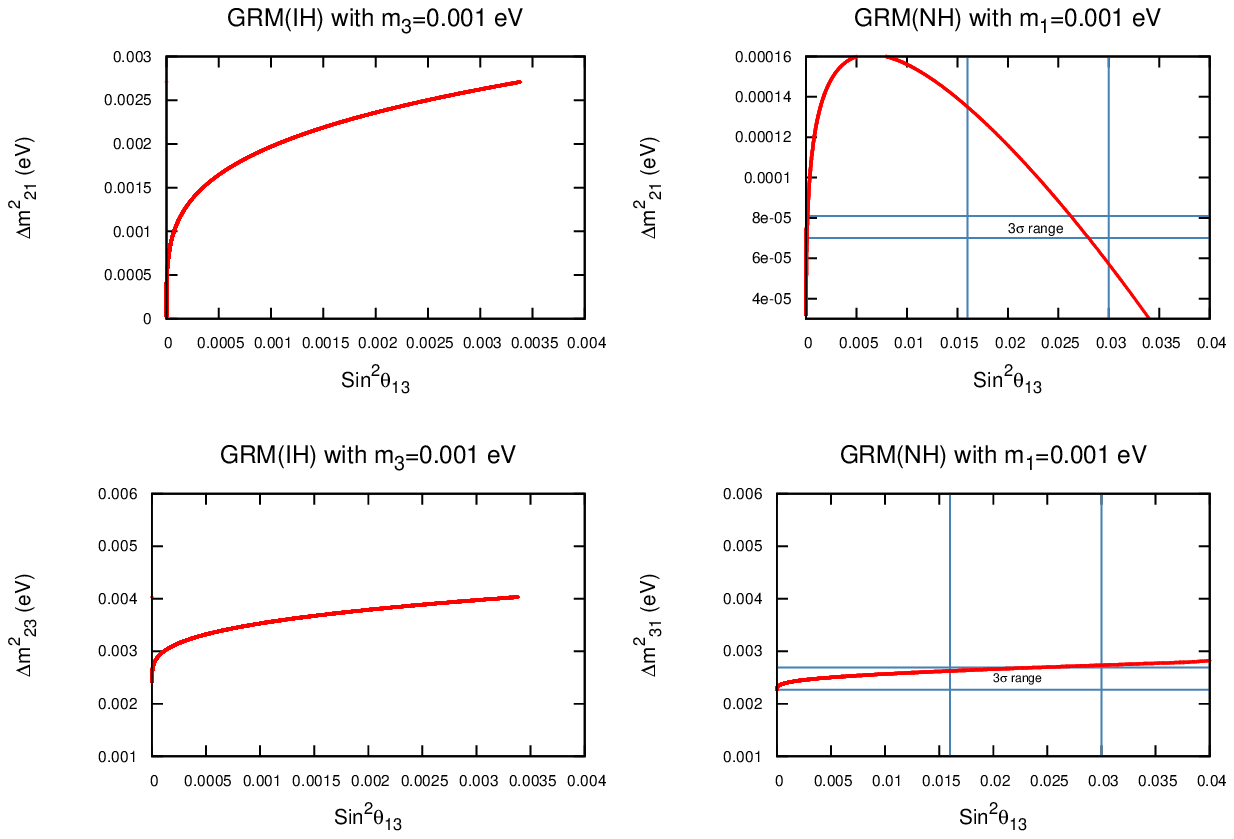}
\caption{$\Delta m^2_{21}$, $\Delta m^2_{23}$, $\Delta m^2_{31}$ with $\sin^2\theta_{13}$ for GRM with $m_1 (m_3) = 0.001$ eV.}
\label{fig:fig17}
\end{figure}

After determining the lepton asymmetry $\epsilon_1$, the corresponding baryon asymmetry can be obtained by
\begin{equation}
Y_B = c \kappa \frac{\epsilon}{g_*}
\end{equation}
through electroweak sphaleron processes \cite{sphaleron}. Here the factor $c$ is measure of the fraction of lepton asymmetry being 
converted into baryon asymmetry and is approximately equal to $-0.55$. $\kappa$ is the dilution factor due to wash-out process which 
erase the produced asymmetry and can be parametrized as \cite{kolbturner}
\begin{eqnarray}
-\kappa &\simeq &  \sqrt{0.1K} \text{exp}[-4/(3(0.1K)^{0.25})], \;\; \text{for} \; K  \ge 10^6 \nonumber \\
&\simeq & \frac{0.3}{K (\ln K)^{0.6}}, \;\; \text{for} \; 10 \le K \le 10^6 \nonumber \\
&\simeq & \frac{1}{2\sqrt{K^2+9}},  \;\; \text{for} \; 0 \le K \le 10.
\end{eqnarray}
where K is given as
$$ K = \frac{\Gamma_1}{H(T=M_1)} = \frac{(m^{\dagger}_{LR}m_{LR})_{11}M_1}{8\pi v^2} \frac{M_{Pl}}{1.66 \sqrt{g_*}M^2_1} $$
Here $\Gamma_1$ is the decay width of $N_1$ and $H(T=M_1)$ is the Hubble constant at temperature $T = M_1$. The factor $g_*$ is the 
effective number of relativistic degrees of freedom at $T=M_1$ and is approximately $110$.

We note that the lepton asymmetry shown in equation (\ref{noflavor}) is obtained by summing over all the flavors $\alpha = e, \mu, \tau$. 
A non-vanishing lepton asymmetry is generated only when the right handed neutrino decay is out of equilibrium. Otherwise both the forward 
and the backward processes will happen at the same rate resulting in a vanishing asymmetry. Departure from equilibrium can be estimated by 
comparing the interaction rate with the expansion rate of the Universe. At very high temperatures $(T \geq 10^{12} \text{GeV})$ all charged 
lepton flavors are out of equilibrium and hence all of them behave similarly resulting in the one flavor regime. However at temperatures 
$ T < 10^{12}$ GeV $(T < 10^9 \text{GeV})$, interactions involving tau (muon) Yukawa couplings enter equilibrium and flavor effects become 
important \cite{flavorlepto}. Taking these flavor effects into account, the final baryon asymmetry is given by 
\begin{equation}
Y^{2 flavor}_B = \frac{-12}{37g^*}[\epsilon_2 \eta\left (\frac{417}{589}\tilde{m_2} \right)+\epsilon^{\tau}_1\eta\left (\frac{390}{589}
\tilde{m_{\tau}}\right )] \nonumber
\end{equation}
\begin{equation}
Y^{3 flavor}_B = \frac{-12}{37g^*}[\epsilon^e_1 \eta\left (\frac{151}{179}\tilde{m_e}\right)+ \epsilon^{\mu}_1 \eta\left (\frac{344}{537}
\tilde{m_{\mu}}\right)+\epsilon^{\tau}_1\eta\left (\frac{344}{537}\tilde{m_{\tau}} \right )] \nonumber
\end{equation}
where $\epsilon_2 = \epsilon^e_1 + \epsilon^{\mu}_1, \tilde{m_2} = \tilde{m_e}+\tilde{m_{\mu}}, \tilde{m_{\alpha}} = \frac{(m^*_{LR})_{\alpha 1} 
(m_{LR})_{\alpha 1}}{M_1}$. The function $\eta$ is given by 
$$ \eta (\tilde{m_{\alpha}}) = \left [\left ( \frac{\tilde{m_{\alpha}}}{8.25 \times 10^{-3} \text{eV}} \right )^{-1}+ \left ( \frac{0.2\times 
10^{-3} \text{eV}}{\tilde{m_{\alpha}}} \right )^{-1.16} \right ]^{-1} $$
In the presence of an additional scalar triplet, the right handed neutrino can also decay through a virtual triplet as shown in figure 
\ref{fig01}. The contribution of this diagram to lepton asymmetry can be estimated as \cite{tripletlepto}
\begin{equation} 
\epsilon^{\alpha}_{\Delta 1}=-\frac{M_1}{8\pi v^2} \frac{\sum_{j=2,3} \text{Im} [(m_{LR})_{1j}(m_{LR})_{1\alpha}(M^{II*}_{\nu})_{j\alpha}]}
{\sum_{j=2,3} \lvert (m_{LR})_{1j}\rvert^2}
\label{eps3}
\end{equation}

For the calculation of baryon asymmetry, we go to the basis where the right handed Majorana neutrino mass matrix is diagonal
\begin{equation}
U^*_R M_{RR} U^{\dagger}_R = \text{diag}(M_1, M_2, M_3)
\label{mrrdiag}
\end{equation}
In this diagonal $M_{RR}$ basis, the Dirac neutrino mass matrix also changes to 
\begin{equation}
m_{LR} = m^0_{LR} U_R
\label{mlrdiag}
\end{equation}
where $m^0_{LR}$ is the Dirac neutrino mass matrix given by
\begin{equation}
m^0_{LR} = U_l m^d_{LR} U^{\dagger}_l
\end{equation}
Here $m^d_{LR}$ is the diagonal form of the Dirac neutrino mass matrix in our calculation given by 
\begin{equation}
m^d_{LR}=\left(\begin{array}{ccc}
\lambda^m & 0 & 0\\
0 & \lambda^n & 0 \\
0 & 0 & 1
\end{array}\right)m_f
\label{mLR1}
\end{equation}
where $\lambda = 0.22$ is the standard Wolfenstein parameter and $(m,n)$ are positive integers. As mentioned earlier, $U_l$ is the matrix 
which is assumed to diagonalize both the charged lepton and Dirac neutrino mass matrices.

\begin{table}
  \centering
  \begin{tabular}{ | l | l |l|l|l|}
    \hline
Parameters  & TBM(IH)& TBM(NH) &BM(IH)&HEX (NH)  \\ \hline
$w$ &0.004435 &0.004575&0.00461&0.00461\\ \hline
$\sin^2\theta_{13}$& 0.01621&0.01672&0.01622&0.01621\\ \hline
$\sin^2\theta_{23}$& 0.4105&0.5918&0.4102&0.5937\\ \hline
  \end{tabular}
  \caption{Parameters used in the calculation of baryogenesis}
  \label{table:wforyb}
\end{table}
\begin{table}
  \centering
  \begin{tabular}{ | l | l | l |}
    \hline
Model & $\delta $ for 1 flavor(in radian)&$\delta $ for 2 flavor(in radian) \\ \hline
TBM(IH), $m_3=0.001$ & 0.00329867-0.0043982297, 3.1376656-3.13860814 & 3.14190681\\ \hline
TBM(NH), $m_1=0.001$& 3.14269221-3.14300637, 6.282085749-6.282242829&-\\ \hline
BM(IH), $m_3=0.001$ & 0.000314159, 1.40711935, 4.8754376&0.0001570769\\ \hline
HEX(NH), $m_1=0.001$ & 3.182276-3.1981413, 6.28020079-6.2808291&- \\ \hline
 
  \end{tabular}
  \caption{Values of $\delta$ giving rise to correct baryon asymmetry}
  \label{table:delta}
\end{table}

\section{Lepton Flavour Violation}
It is known that neutrino flavor is violated in neutrino sector from the experimental observed oscillation phenomena.  
In the previous sections, we have presented the model for explaining the non-zero values of reactor mixing angle 
$\theta_{13}$, as revealed from recent oscillation experiment, by perturbation method. The idea is to take the type I 
seesaw contribution as the leading term in the neutrino mass matrix with the various choices for mixing matrices such 
as Bimaximal, Tribimaximal, Hexagonal and Golden ratio type with uniquely predictlng $\theta_{13}=0$. In the second step, 
the non-zero value for reactor mixing angle can be found by adding a small perturbation matrix (type II seesaw term 
in the present work) parametrized by $\omega$ to the leading order mass matrix (type I seesaw term). It is numerically 
examined that the overall scale of the perturbation $\omega$ is derived to be order of $0.001$ eV. 

The analytic expression for Higgs triplet VEV generated from the trilinear mass term $\mu_{\Phi \Delta}$ in the scalar 
Lagrangian given in eq.(\ref{eq:scalar_lag}) is given below
\begin{eqnarray}
v_\Delta = \mu_{\Phi \Delta} v^2 {\Large /} \sqrt{2} M^2
\end{eqnarray}
With $v^2=v^2+v_\Delta^2\simeq \mbox{(174\, GeV)}^2$. The resulting mass fomrula for neutrino mass 
is 
$$m_\nu = f_\nu v_\Delta \, .$$

The numerical value of perturbation term $\omega \simeq f_\nu v_\Delta$ is crucially depend upon the Majorana coupling $f_\nu$, 
trilinear mass parameter $\mu_{\Phi \Delta}$ and $M$ which has fixed around TeV scale. Thus, the required 
value of $\omega=0.001$ eV can be easily obtained by choosing these parameters appropriately. We intend to examine 
possible Lepton Flavour Violation (LFV) within the present framework with TeV scale scalar triplet. 

\begin{figure}[t!]
\centering
\includegraphics[scale=0.58,angle=0]{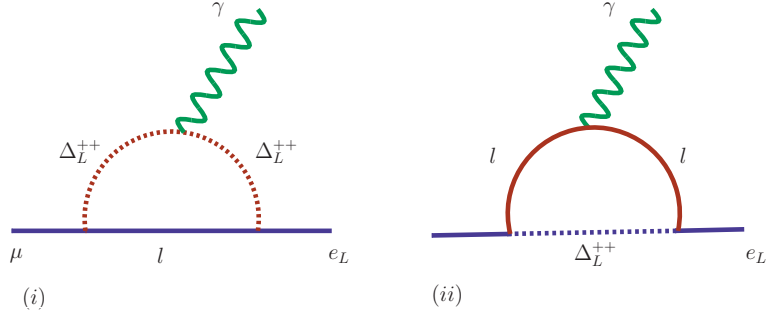}
\caption{Relevant Feynman diagram for inverse neutrinoless double beta decay 
         due to light neutrino exchange diagram.}
\label{Feyn:mu-egamma}
\end{figure}
\begin{figure}[h!]
\centering
\includegraphics[scale=0.78,angle=0]{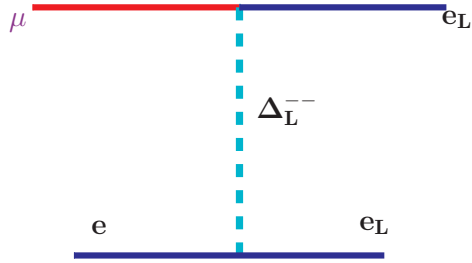}
\caption{Relevant Feynman diagram for inverse neutrinoless double beta decay 
         due to light neutrino exchange diagram.}
\label{Feyn:mu-3e}
\end{figure}

Within the present framework having type-I + type II seesaw mechanism having TeV scale Higgs triplet and heavy RH Majorana neutrinos, 
there are various Feynman diagrams contributing to the LFV processes like $\mu \to e \gamma$, $\mu\to eee$ and $\mu \to e$ conversion 
inside a nuclei. They are arising from exchange of: (i) light neutrino exchange, (ii) heavy neutrinos through light-heavy neutrino mixing 
propertional to $M_D/M_R$, (iii) charged Higgs scalars. Since the light neutrino contribution is heavily suppressed ($\mbox{Br.}(\mu \to e 
\gamma) \simeq 10^{-50}$) and heavy neutrino contribution is very much suppressed as well due to large mass for RH Majorana neutirno $M_R > 
10^{9}$ GeV. Thus, the Higgs triplet scalar with TeV mass range can contribute to various LFV processes as shown in Fig.\ref{Feyn:mu-egamma} 
for $\mu \to e \gamma$ and Fig.\ref{Feyn:mu-3e} for $\mu \to eee$.

\begin{eqnarray}
\text{Br}_{\mu \to e + \gamma}^{\rm triplet} \simeq 0.01 \times \frac{M^4_{W_L}}{g^4_L} 
                  \bigg| \frac{(f^* f)_{21}}{M^2_{\Delta^{++}_L}}\bigg|^2 
\simeq 0.01 \times \frac{v^4}{v^4_{\Delta}}
              \bigg|\frac{(m^{II}_\nu\, {m^{II}_{\nu}}^\dagger)_{12}}{M^2_{\Delta^{++}_L}}\bigg|^2\, ,
\end{eqnarray}
where we have used $M_{W_L} \simeq 1/2 g_L v$, $m^{II}_\nu \equiv \omega \simeq {\cal O}(0.001)$ eV. 
Using $v_{\Delta} \simeq 10^{-5} - 10^{-9}$ GeV, $v \simeq 174$ GeV and doubly charged Higgs scalar mass 
around 300 GeV, the model prediction for $\text{Br}_{\mu \to e + \gamma}^{\rm triplet}$ is found to be 
$\text{Br}\left(\mu \rightarrow e + \gamma \right)|_{\rm theory} < 4.2 \times 10^{-15}$. 
The current experimental constraint on $\mu \to e \gamma$ is 
$$\text{Br}\left(\mu \rightarrow e + \gamma \right)\big|_{\rm expt.} < 2.4 \times 10^{-12}$$ at 
90\% C.L \cite{meg} 
Although the predicted value of branching ratio for LFV decays $\mu \to e \gamma$ is beyond the reach 
of current experimental sensitivity (for more details, see Refs.\cite{lfv-typeII,lfv-typeII-a}), it is 
hoped to be probed at future experiment \cite{baldini} which is planned to reach $\text{Br}\left(\mu 
\rightarrow e + \gamma \right)|_{\rm future} \simeq \times 10^{-16}$. 

%
Another important LFV decays $\mu \to 3e$ shown in Fig.\ref{Feyn:mu-3e} for which 
the branching ratio is found to be
\begin{eqnarray}
\text{Br}_{\mu \rightarrow 3e}^{\rm triplet} \simeq \frac{1}{2} \frac{M^4_{W_L}}{g^4_L} 
              \bigg|\frac{(f f^\dagger)_{12}}{M^2_{\Delta^{++}_L}}\bigg|^2 
\simeq \frac{1}{2} \frac{v^4}{v^4_{\Delta}}
              \bigg|\frac{(m^{II}_\nu\, {m^{II}_{\nu}}^\dagger)_{12}}{M^2_{\Delta^{++}_L}}\bigg|^2\, . 
\end{eqnarray}
For numerical evaluation, consider $v_\Delta \simeq $ eV, $f\simeq 10^{-3.5}$, and $M_{\Delta^{++}_{L}} 
\simeq (300-1000)$ GeV results $\text{Br}\left(\mu \rightarrow 3e \right) \simeq 10^{-12}-10^{-16}$ 
which is accessible to the ongoing or future planned search experiments. The present experimental 
upper bound for $\mu \to 3e$ process is $\mbox{Br.}(\mu \to 3e) < 1.0 \times 10^{-12}$ \cite{sindrum} 
while it is planned to reach $\mbox{Br.}(\mu \to 3e) < 10^{-16}$ \cite{sind-prop}.


\section{Numerical Analysis}
\label{sec:numeric}
To begin with, we write down the light neutrino mass matrix $m_{LL}$ in terms of (complex) mass eigenvalues 
$m_1$, $m_2$, $m_3$ and PMNS mixing matrix $U_{\rm PMNS} \equiv U$, working in a basis where charged lepton 
mass matrix is already diagonal, as
\begin{eqnarray}
m_\nu=U^* \mbox{diag}(m_1, m_2, m_3) U^\dagger\, .
\end{eqnarray}
The mixing matrix $U$ is parametrized in terms of three neutrino mixing angle $\theta_{23}$, $\theta_{12}$, 
$\theta_{13}$ and a Dirac phase $\delta$. The two Majorana phases are absorbed in mass eigenvalues $m_i$ instead 
in the mixing matrix $U$. Here the two Majorana phases are simply taken to be zero for subsequent numerical analysis.

At the first step of numerical analysis, we have considered particularly four choices of $U$ and $m_\nu$ so that 
$\theta_{13}=0$ and $\theta_{23}$ is maximal with the general form of the mixing matrix at leading order as
\begin{eqnarray}
U=\begin{pmatrix}
   c_{12}            & s_{12}             &  0             \\
  -s_{12}/\sqrt{2}   & c_{12}/\sqrt{2}    & -1/\sqrt{2}    \\
  -s_{12}/\sqrt{2}   & c_{12}/\sqrt{2}    &  1/\sqrt{2}    
  \end{pmatrix}
\end{eqnarray}
We start writing the relevant matrix form for light neutrino mass satisfying $\mu-\tau$ symmetry 
and corresponding mixing matrix having different values of $\theta_{12}$ but consistent with our 
earlier assumptions, i.e. $\theta_{13}=0$ and maximal $\theta_{23}$, as
\begin{eqnarray}
m^{(0)}_{\nu}\big|_{\rm BM} 
= \begin{pmatrix}
 A+B&F&F\\
 F&A&B\\
 F&B&A
 \end{pmatrix} \, , 
 \quad \quad U_{\rm BM}=\begin{pmatrix}
   1/\sqrt{2}            & 1/\sqrt{2}             &  0             \\
  -1/2   & 1/2    & -1/\sqrt{2}    \\
  -1/2   & 1/2    &  1/\sqrt{2}    
  \end{pmatrix}\, ,
\end{eqnarray}
with $m_1=A+B+\sqrt{2}F, m_2=A+B-\sqrt{2}F, m_3=A-B$. 


\begin{eqnarray}
m^{(0)}_{\nu}\big|_{\rm TBM} 
= \begin{pmatrix}
  A& B&B\\
B& A+F & B-F \\
B & B-F & A+F 
  \end{pmatrix}
 \, , 
 \quad \quad U_{\rm TBM}=\begin{pmatrix}
   2/\sqrt{6}    & 1/\sqrt{3}             &  0             \\
  -1/\sqrt{6}    & 1/\sqrt{3}    & -1/\sqrt{2}    \\
  -1/\sqrt{6}    & 1/\sqrt{3}    &  1/\sqrt{2}    
  \end{pmatrix}\, , 
\end{eqnarray}
with $m_1 = A-B, \; m_2 = A+2B, \; m_3 = A-B+2F$. It is clear from the BM and TBM-type of 
mixing matrices 
$$\tan^2 \theta_{23} = |U_{\mu 3}|^2/|U_{\tau 3}|^2 =1\,.$$

Similarly, there are other two other types of mass matrix and mixing matrix which can reproduce 
$\theta_{13}=0$ and maximal $\theta_{23}$ and they are: (i) Hexagonal type predicting $\theta_{12} 
=\pi/6$, (ii) Golden ratio type for which $\theta_{12} \tan^{-1}(1/\varphi)$ with $\phi=(1+\sqrt{5})/2$. 
\begin{eqnarray}
m^{(0)}_{\nu}\big|_{\rm HM} 
= \begin{pmatrix}
 A&B&B\\
 B&\frac{1}{2}(A+2\sqrt{\frac{2}{3}}B+F)&\frac{1}{2}(A+2\sqrt{\frac{2}{3}}B-F)\\
 B&\frac{1}{2}(A+2\sqrt{\frac{2}{3}}B-F)&\frac{1}{2}(A+2\sqrt{\frac{2}{3}}B+F)
  \end{pmatrix}
 \, , 
 U_{\rm HM}=\begin{pmatrix}
  \frac{\sqrt{3}}{2}   & \frac{1}{2}         & 0   \\
  -\frac{\sqrt{2}}{4}  & \frac{\sqrt{6}}{4}  & -\frac{1}{\sqrt{2}} \\
  -\frac{\sqrt{2}}{4}  & -\frac{\sqrt{6}}{4} &  \frac{1}{\sqrt{2}}
  \end{pmatrix}\, \nonumber
\end{eqnarray}
with $m_1=\frac{1}{3}(3A-\sqrt{6}B), m_2=A+\sqrt{6}B$ and $m_3=F$.


\begin{eqnarray}
m^{(0)}_{\nu}\big|_{\rm GRM} 
= \begin{pmatrix}
 A&B&B\\
 B&F&A+\sqrt{2}B-F\\
 B&A+\sqrt{2}B-F&F
  \end{pmatrix}
 \, , 
 U_{\rm GRM}=\begin{pmatrix}
  \frac{\sqrt{2}}{\sqrt{5-\sqrt{5}}}    &  \frac{\sqrt{2}}{\sqrt{5+\sqrt{5}}}    &  0    \\
  -\frac{\sqrt{2}}{\sqrt{5+\sqrt{5}}}   & \frac{\sqrt{2}}{\sqrt{5-\sqrt{5}}}    & -1/\sqrt{2}    \\
   -\frac{\sqrt{2}}{\sqrt{5+\sqrt{5}}}  & \frac{\sqrt{2}}{\sqrt{5-\sqrt{5}}}    &  1/\sqrt{2}    
  \end{pmatrix}\, \nonumber
\end{eqnarray}
with $m_1=\frac{1}{2}(2A+\sqrt{2}B-\sqrt{10}B), m_2=\frac{1}{2}(2A+\sqrt{2}B+\sqrt{10}B),$ 
and $m_3=-A-\sqrt{2}B+2F$. 

For normal hierarchy, the diagonal mass matrix of the light neutrinos can be written as $m_{\text{diag}} 
= \text{diag}(m_1, \sqrt{m^2_1+\Delta m_{21}^2}, \sqrt{m_1^2+\Delta m_{31}^2})$ whereas for inverted 
hierarchy it can be written as $m_{\text{diag}} = \text{diag}(\sqrt{m_3^2+\Delta m_{23}^2-\Delta m_{21}^2}, 
\sqrt{m_3^2+\Delta m_{23}^2}, m_3)$.  We choose two possible values of the lightest mass eigenstate $m_1, m_3$ 
for normal and inverted hierarchies respectively. First we choose $m_{\text{lightest}}$ as large as possible such 
that the sum of the absolute neutrino masses fall just below the cosmological upper bound. For normal and inverted 
hierarchies, this turns out to be $0.07$ eV and $0.065$ eV respectively. Then we allow moderate hierarchy to exist 
between the mass eigenvalues and choose the lightest mass eigenvalue to be $0.001$ eV to study the possible changes 
in our analysis and results. The parametrization for all these possible cases are shown in table\, \ref{table:BM}, 
\ref{table:TBM}, \ref{table:HM} and \ref{table:GRM}.

For our numerical analysis, we adopt the minimal structure (\ref{matrixt2}) of the type II seesaw term 
as 
\begin{equation}
m^{II}_{LL}=\left(\begin{array}{ccc}
0& -w& w\\
-w& w & 0 \\
w & 0& -w
\end{array}\right)\, ,
\label{matrix3}
\end{equation}
where $\omega$ denotes the strength of perturbation coming from type II seesaw 
mechanism.


We first numerically fit the leading order $\mu-\tau$ symmetric neutrino mass matrix (\ref{matrix1}) by 
taking the central values of the global fit neutrino oscillation data \cite{schwetz12}. We also incorporate 
the cosmological upper bound on the sum of absolute neutrino masses \cite{Planck13} reported by the Planck 
collaboration recently. In the second step, we have to diagonalize the complete mass matrix
$$m_{\nu}=m^{(0)}_\nu + m^{\rm (pert.)}_\nu = m^{I}_\nu + m^{II}_{\nu}\, , $$
and as a result, there is a corresponding mixing matrix whose elements are related to the parameters of 
the model plus the strength of the type II perturbation term.

\begin{figure}
\centering
\includegraphics[width=0.85\textwidth]{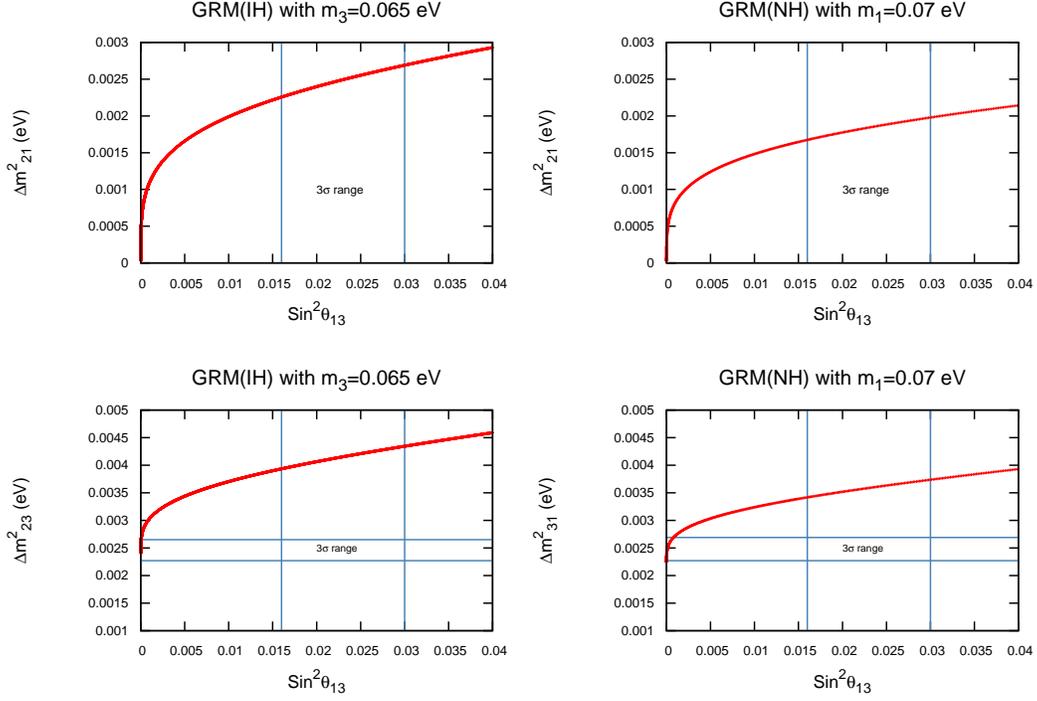}
\caption{$\Delta m^2_{21}$, $\Delta m^2_{23}$, $\Delta m^2_{31}$ with $\sin^2\theta_{13}$ for GRM with $m_1 (m_3) = 0.07 (0.065)$ eV.}
\label{fig:fig18}
\end{figure}
\begin{figure}
\centering
\includegraphics[width=0.85\textwidth]{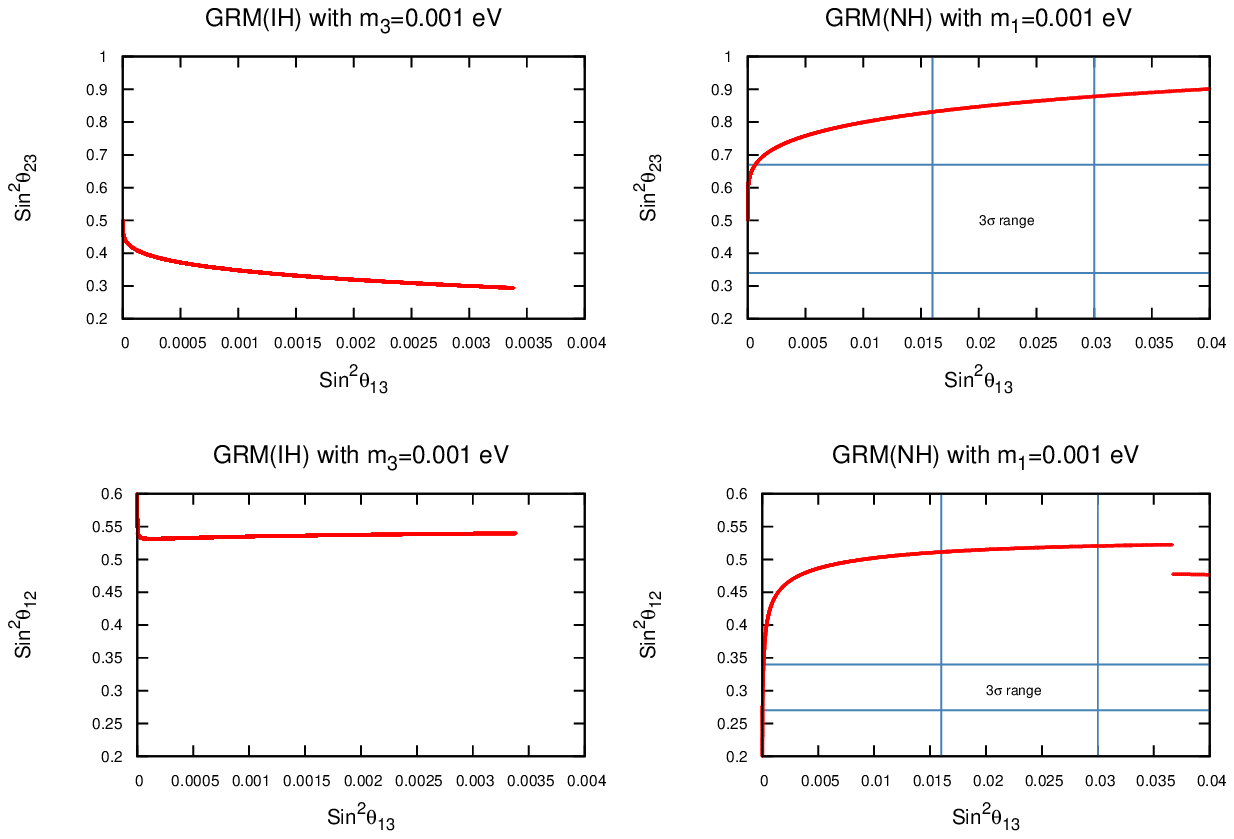}
\caption{$\sin^2\theta_{23}$, $\sin^2\theta_{12}$ with $\sin^2\theta_{13}$ for GRM with $m_1 (m_3) = 0.001$ eV.}
\label{fig:fig19}
\end{figure}
\begin{figure}
\centering
\includegraphics[width=0.85\textwidth]{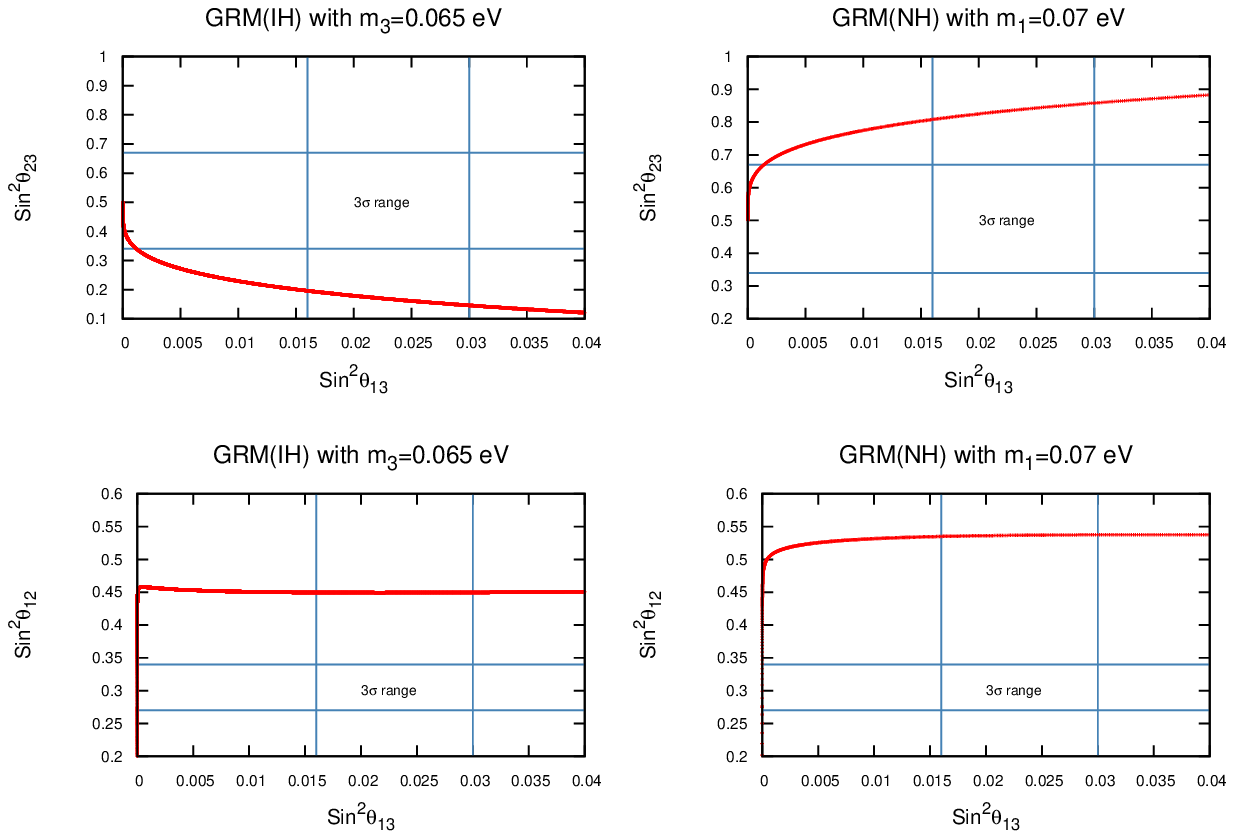}
\caption{$\sin^2\theta_{23}$, $\sin^2\theta_{12}$ with $\sin^2\theta_{13}$ for GRM with $m_1 (m_3) = 0.07 (0.065)$ eV.}
\label{fig:fig20}
\end{figure}
\begin{figure}
\centering
\includegraphics[width=0.85\textwidth]{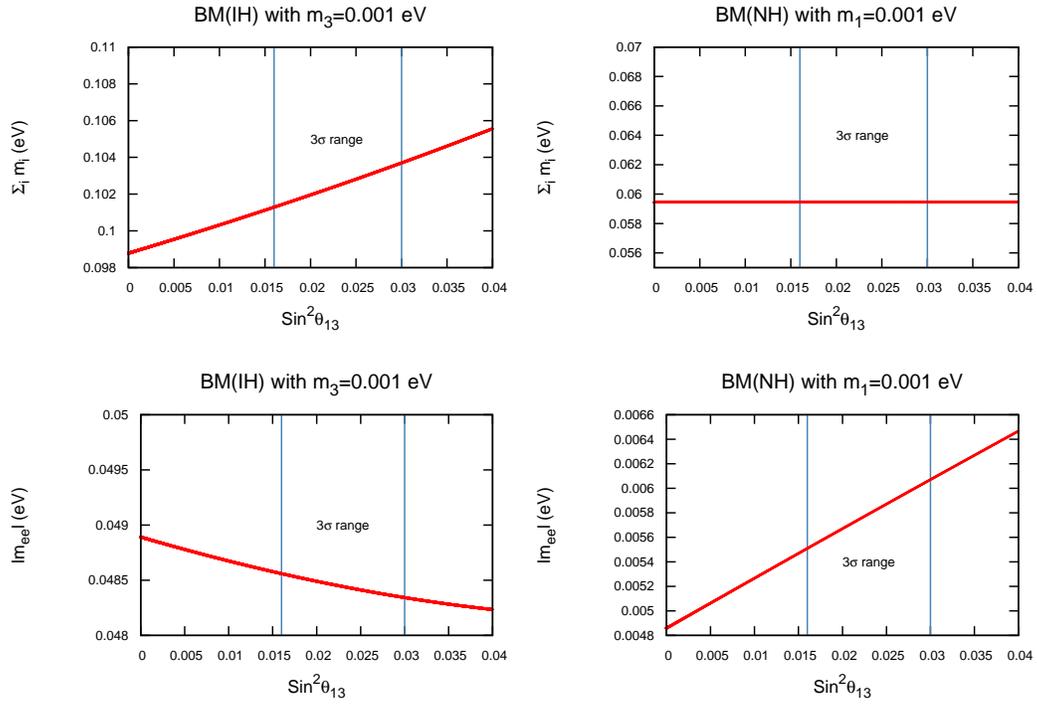}
\caption{$\sum_i \lvert m_i \rvert$, $\lvert m_{ee} \rvert$ with $\sin^2\theta_{13}$ for BM with $m_1 (m_3) = 0.001$ eV.}
\label{fig:fig21}
\end{figure}
\begin{figure}
\centering
\includegraphics[width=0.85\textwidth]{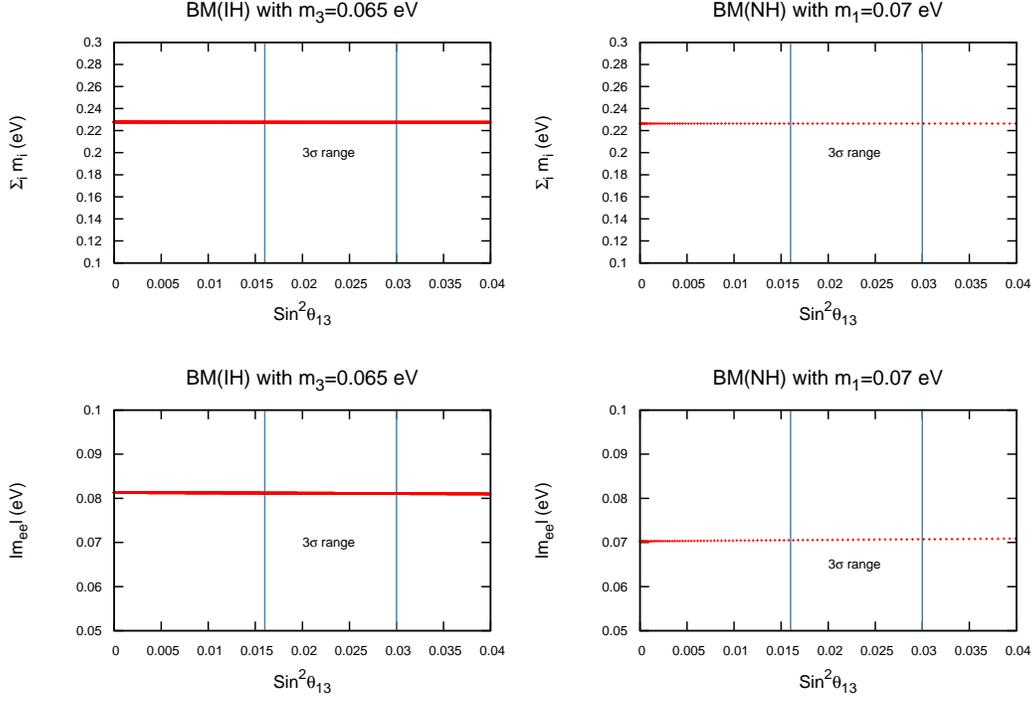}
\caption{$\sum_i \lvert m_i \rvert$, $\lvert m_{ee} \rvert$ with $\sin^2\theta_{13}$ for BM with $m_1 (m_3) = 0.07 (0.065)$ eV. }
\label{fig:fig22}
\end{figure}
\begin{figure}
\centering
\includegraphics[width=0.85\textwidth]{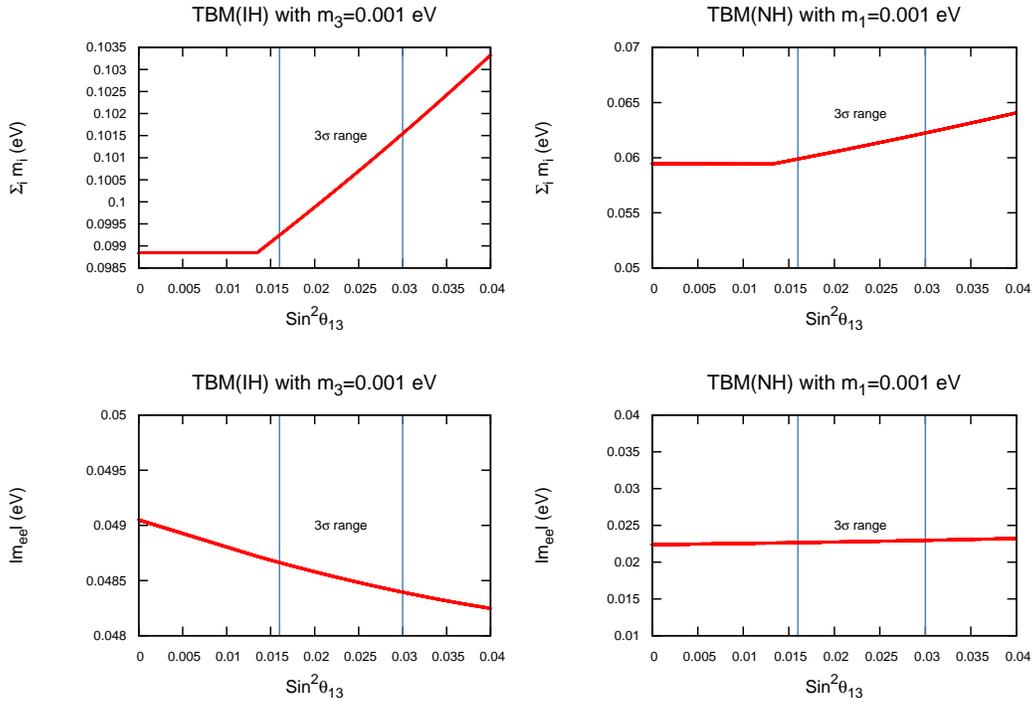}
\caption{$\sum_i \lvert m_i \rvert$, $\lvert m_{ee} \rvert$ with $\sin^2\theta_{13}$ for TBM with $m_1 (m_3) = 0.001$ eV.}
\label{fig:fig23}
\end{figure}
\begin{figure}
\centering
\includegraphics[width=0.85\textwidth]{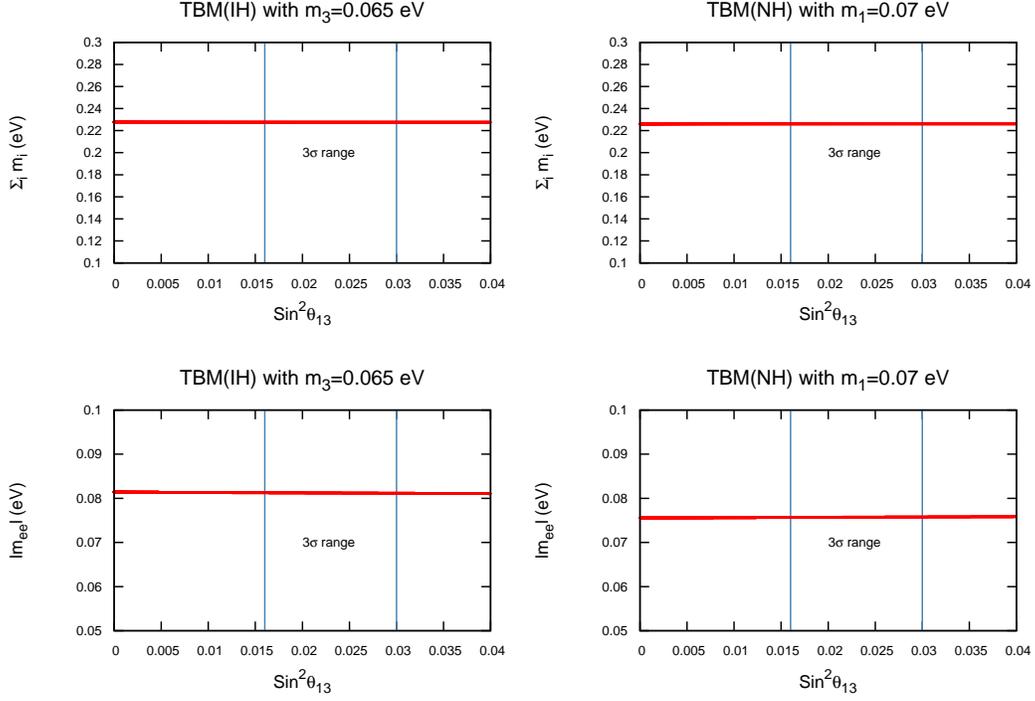}
\caption{$\sum_i \lvert m_i \rvert$, $\lvert m_{ee} \rvert$ with $\sin^2\theta_{13}$ for TBM with $m_1 (m_3) = 0.07 (0.065)$ eV.}
\label{fig:fig24}
\end{figure}
\begin{figure}
\centering
\includegraphics[width=0.85\textwidth]{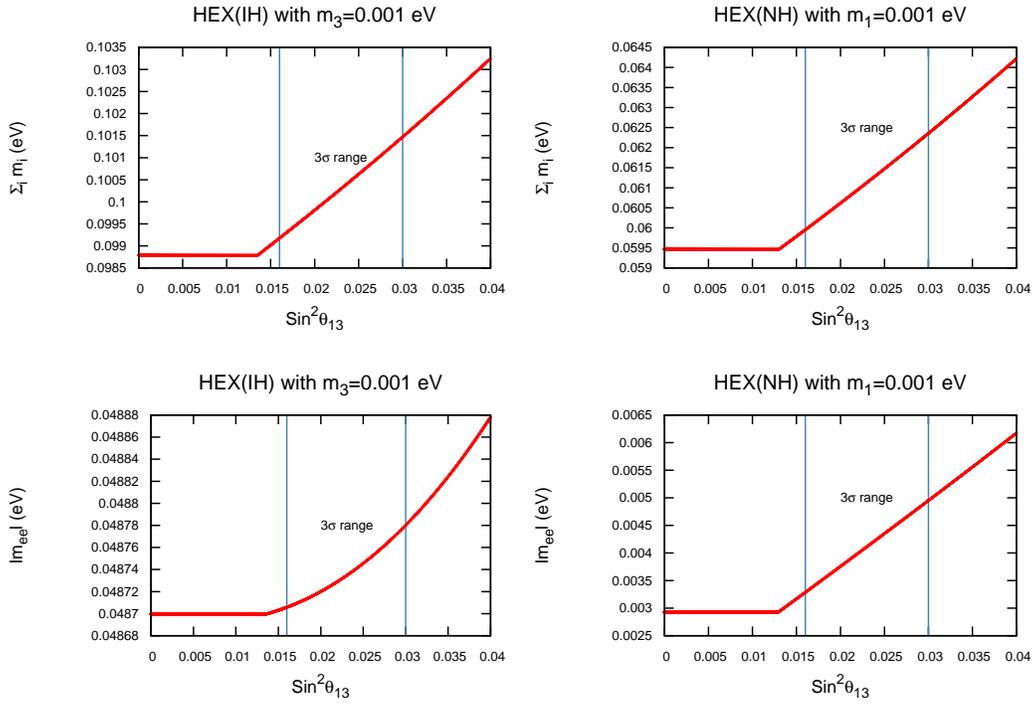}
\caption{$\sum_i \lvert m_i \rvert$, $\lvert m_{ee} \rvert$ with $\sin^2\theta_{13}$ for HM with $m_1 (m_3) = 0.001$ eV. }
\label{fig:fig25}
\end{figure}
\begin{figure}
\centering
\includegraphics[width=0.85\textwidth]{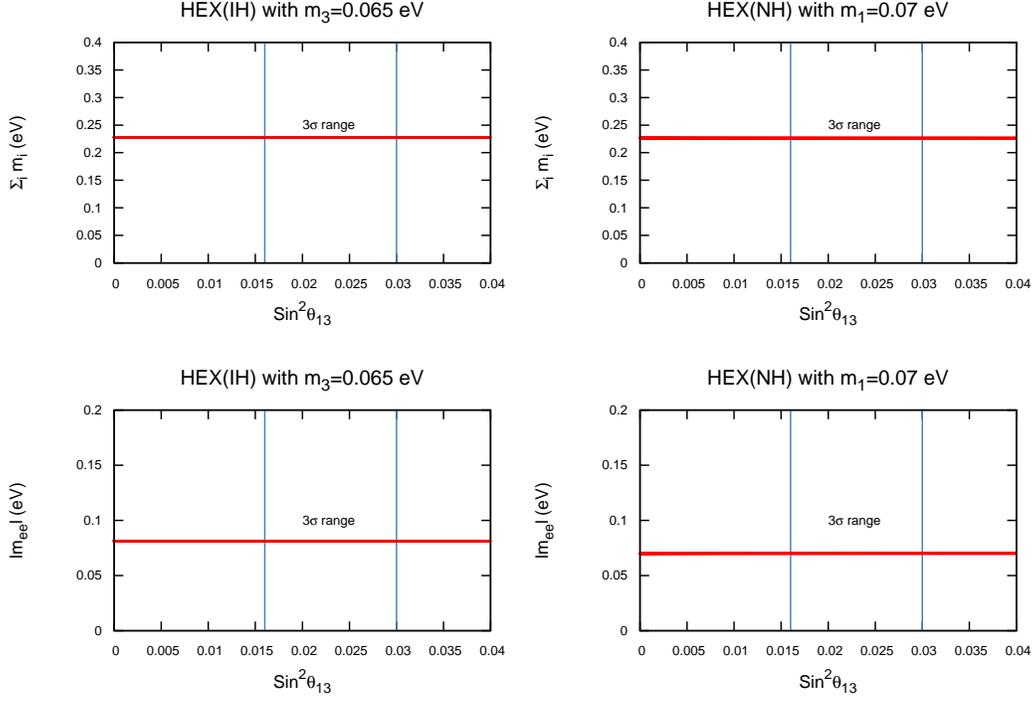}
\caption{$\sum_i \lvert m_i \rvert$, $\lvert m_{ee} \rvert$ with $\sin^2\theta_{13}$ for HM with $m_1 (m_3) = 0.07 (0.065)$ eV.}
\label{fig:fig26}
\end{figure}
\begin{figure}
\centering
\includegraphics[width=0.85\textwidth]{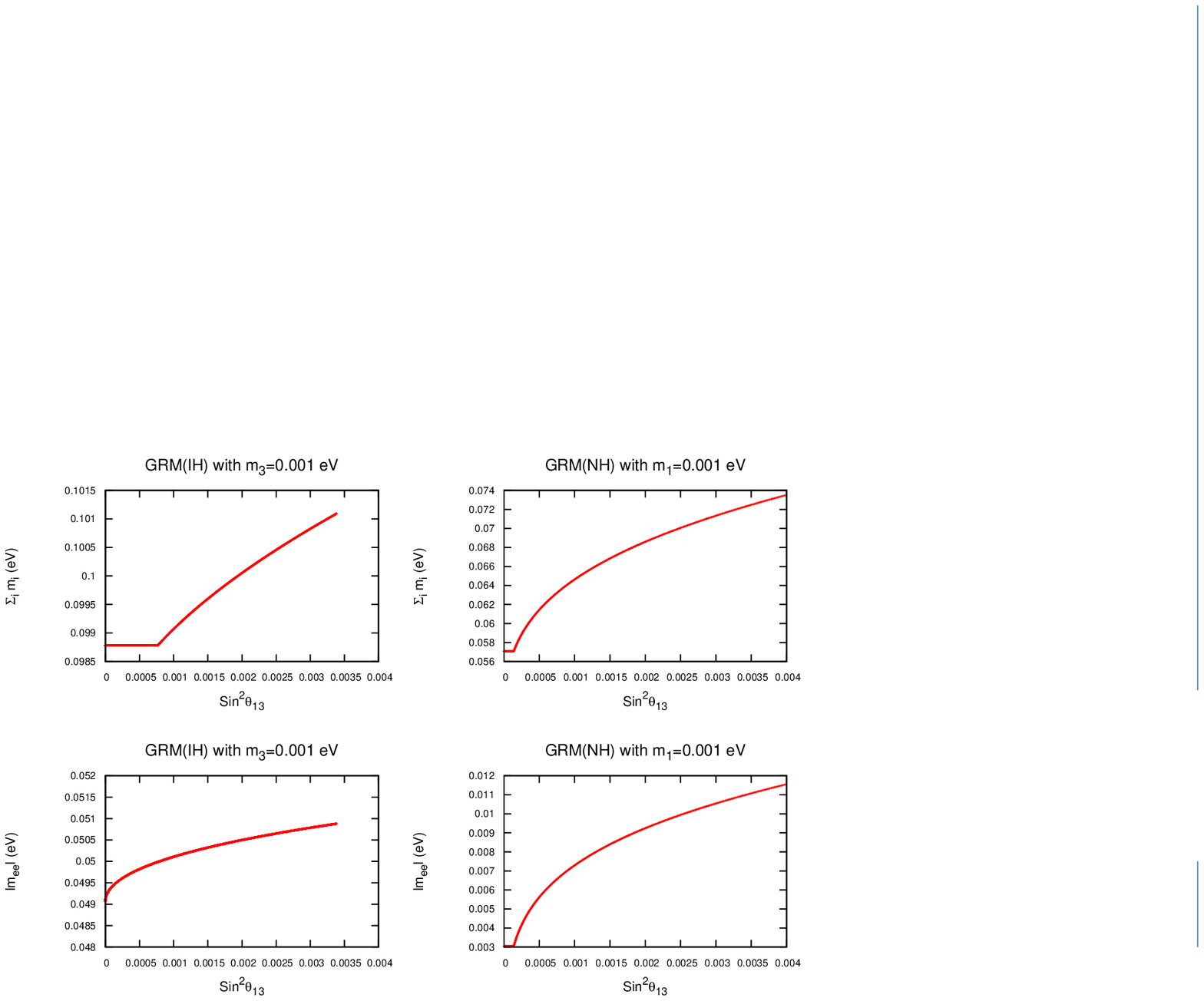}
\caption{$\sum_i \lvert m_i \rvert$, $\lvert m_{ee} \rvert$ with $\sin^2\theta_{13}$ for GRM with $m_1 (m_3) = 0.001$ eV. }
\label{fig:fig27}
\end{figure}
\begin{figure}
\centering
\includegraphics[width=0.85\textwidth]{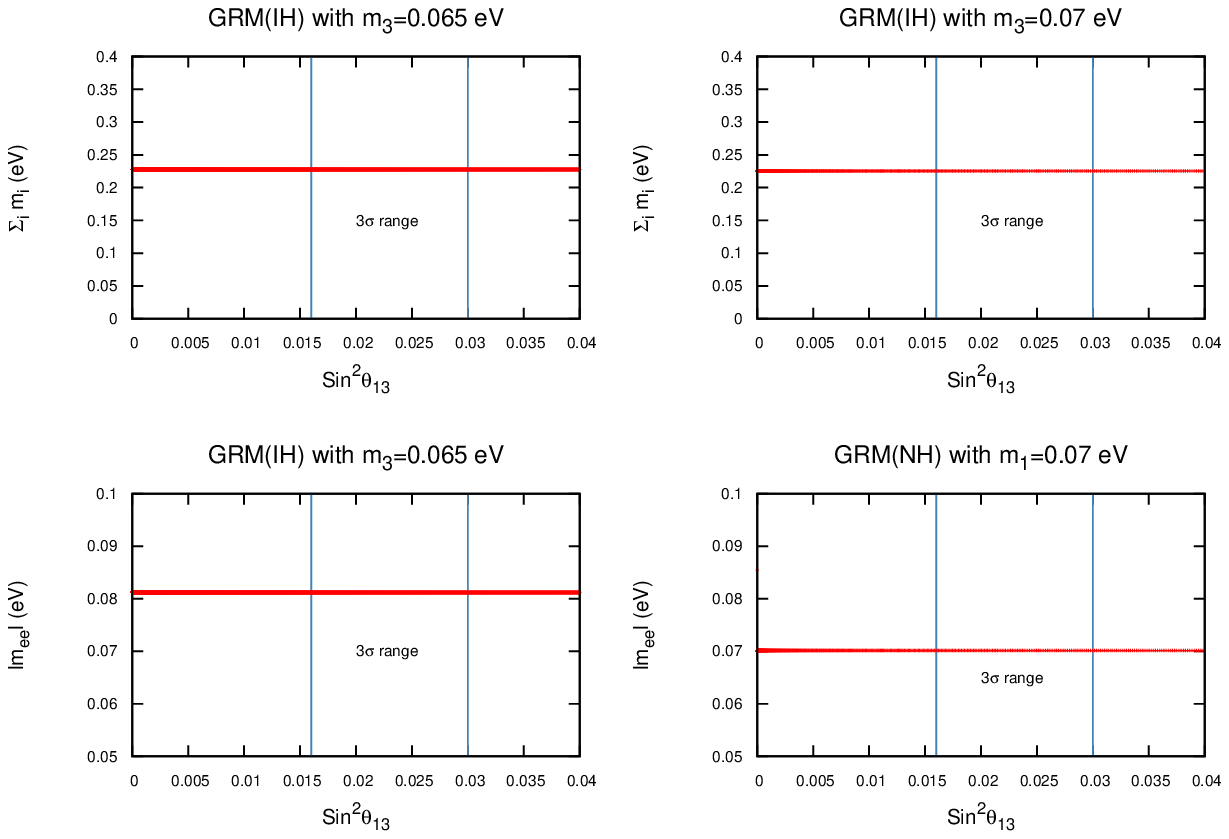}
\caption{$\sum_i \lvert m_i \rvert$, $\lvert m_{ee} \rvert$ with $\sin^2\theta_{13}$ for GRM with $m_1 (m_3) = 0.07 (0.065)$ eV.}
\label{fig:fig28}
\end{figure}

After fitting the type I seesaw contribution to neutrino mass with experimental data, we introduce the type II seesaw 
contribution as a perturbation to the $\mu-\tau$ symmetric neutrino mass matrix. The strength of the type II seesaw perturbation 
in order to generate the correct value of non-zero $\theta_{13}$ can be seen from figure \ref{fig:fig1}, \ref{fig:fig2}, \ref{fig:fig3} 
and \ref{fig:fig4}. We also calculate other neutrino parameters by varying the type II seesaw strength and show our results 
as a function of $\sin^2{\theta_{13}}$ in figure \ref{fig:fig5}, \ref{fig:fig6}, \ref{fig:fig7}, \ref{fig:fig8} for BM mixing, 
figure \ref{fig:fig9}, \ref{fig:fig10}, \ref{fig:fig11}, \ref{fig:fig12} for TBM mixing, figure \ref{fig:fig13}, \ref{fig:fig14}, 
\ref{fig:fig15}, \ref{fig:fig16} for Hexagonal mixing and figure \ref{fig:fig17}, \ref{fig:fig18}, \ref{fig:fig19}, \ref{fig:fig20} 
for GR mixing. We also calculate the sum  of the absolute neutrino masses $\sum_i \lvert m_i\rvert$ to check whether it lies below 
the Planck upper bound. Finally, we calculate the effective neutrino mass $m_{ee}=\lvert \sum_i U_{ei}^2 m_i \rvert$ which can
 play a great role in neutrino-less double beta decay. These are shown as a function of $\sin^2{\theta_{13}}$ in figure \ref{fig:fig21}, 
 \ref{fig:fig22}, \ref{fig:fig23}, \ref{fig:fig24}, \ref{fig:fig25}, \ref{fig:fig26}, \ref{fig:fig27} and \ref{fig:fig28}.

%
 

To calculate the baryon asymmetry, we first have to make a choice of the diagonal Dirac neutrino mass matrix $m^d_{LR}$. The most natural choice is to take $m^d_{LR}$ to be the same as the diagonal charged lepton mass matrix. We check this particular case and find that this corresponds to a three flavor leptogenesis scenario and for all values of Dirac CP phase $\delta$, the resulting baryon asymmetry falls far outside the observed range. As discussed in our earlier work \cite{leptodborah}, here also we assume the parametric form of $m^d_{LR}$ as given in (\ref{mLR1}) and choose the integers $(m,n)$ in such a way that the lightest right handed neutrino mass falls either in one flavor or two flavor or three flavor regime. We take $m_f = 82.43$ GeV and find that the choice $(m,n)=(1,1)$ keeps the lightest right handed neutrino in the one flavor regime, that is $M_1 > 10^{12}$ GeV. For $(m,n) = (3,1)$, the lightest right handed neutrino mass is in the range $10^9 \; \text{GeV} < M_1 < 10^{12} \; \text{GeV}$ 
which corresponds to the two flavor regime of leptogenesis as discussed in the previous section. Similarly, to keep the lightest right handed neutrino mass below $10^9$ GeV, the three flavor regime, we take $(m,n)$ as $(5,3)$ in the Dirac neutrino mass matrix (\ref{mLR1}). The values of type II seesaw strength $w$ and corresponding neutrino neutrino mixing angles used in the calculation of leptogenesis are shown in table \ref{table:wforyb}. The results for final baryon asymmetry are shown in figure \ref{fig:fig29} and \ref{fig:fig30} for BM, TBM and HM models. Since GRM models do not give rise to correct neutrino parameters, we do not calculate the baryon asymmetry for that case.

\begin{table}
  \centering
  \begin{tabular}{ | l | l |l|l|l|l|l|l|l|l|}
    \hline
Model &$\Delta m_{21}^2$&$\shortstack {$\Delta m_{23}^2$\\ $\Delta m_{31}^2$}$& $\theta_{13}$&$\theta_{23}$& $\theta_{12}$ 
& $\sum \lvert m_{i}\rvert$& $\shortstack {$Y_B$ \\\text{(1\, flav.)}}$&$ \shortstack {$Y_B$ \\ \text{(2\, flav.)}}$&$\shortstack {$Y_B$ \\ 
\text{(3\, flav.)}}$  \\ 
\hline \hline
BM(IH) ($m_3$=0.001)&$\checkmark$&$\checkmark$& $\checkmark$&$\checkmark$& $\checkmark$& $\checkmark$& $\checkmark$&$\checkmark$& $\times$  \\ \hline
BM(NH) ($m_1$=0.001)&$\checkmark$&$\checkmark$& $\checkmark$&$\checkmark$& $\times$& $\checkmark$& $\times$&$\times$& $\times$  \\ \hline
BM(IH) ($m_3$=0.065)&$\times$&$\checkmark$& $\checkmark$&$\checkmark$& $\times$& $\checkmark$& $\times$&$\times$& $\times$  \\ \hline
BM(NH) ($m_1$=0.07)&$\checkmark$&$\checkmark$& $\checkmark$&$\checkmark$& $\times$& $\checkmark$& $\times$&$\times$& $\times$  \\ \hline
TBM(IH) ($m_3$=0.001)&$\checkmark$&$\checkmark$& $\checkmark$&$\checkmark$& $\checkmark$& $\checkmark$& $\checkmark$&$\checkmark$& $\times$  \\ \hline
TBM(NH) ($m_1$=0.001)&$\checkmark$&$\checkmark$& $\checkmark$&$\checkmark$& $\checkmark$& $\checkmark$& $\checkmark$&$\times$& $\times$  \\ \hline
TBM(IH) ($m_3$=0.065)&$\times$&$\checkmark$& $\checkmark$&$\checkmark$& $\checkmark$& $\checkmark$& $\times$&$\times$& $\times$  \\ \hline
TBM(NH) ($m_1$=0.07)&$\times$&$\checkmark$& $\checkmark$&$\checkmark$& $\checkmark$& $\checkmark$& $\times$&$\times$& $\times$  \\ \hline
HEX(IH) ($m_3$=0.001)&$\checkmark$&$\checkmark$& $\checkmark$&$\checkmark$& $\times$& $\checkmark$& $\times$&$\times$& $\times$  \\ \hline
HEX(NH) ($m_1$=0.001)&$\checkmark$&$\checkmark$& $\checkmark$&$\checkmark$& $\checkmark$& $\checkmark$& $\checkmark$&$\times$& $\times$  \\ \hline
HEX(IH) ($m_3$=0.065)&$\times$&$\checkmark$& $\checkmark$&$\checkmark$& $\checkmark$& $\checkmark$& $\times$&$\times$& $\times$  \\ \hline
HEX(NH) ($m_1$=0.07)&$\times$&$\checkmark$& $\checkmark$&$\checkmark$& $\checkmark$& $\checkmark$& $\times$&$\times$& $\times$  \\ \hline
GRM(IH) ($m_3$=0.001)&$\times$&$\times$& $\times$&$\times$& $\times$& $\checkmark$& $\times$&$\times$& $\times$  \\ \hline
GRM(NH) ($m_1$=0.001)&$\checkmark$&$\checkmark$& $\checkmark$&$\times$& $\times$& $\checkmark$& $\times$&$\times$& $\times$  \\ \hline
GRM(IH) ($m_3$=0.065)&$\times$&$\times$& $\checkmark$&$\times$& $\times$& $\checkmark$& $\times$&$\times$& $\times$  \\ \hline
GRM(NH) ($m_1$=0.07)&$\times$&$\times$& $\checkmark$&$\times$& $\times$& $\checkmark$& $\times$&$\times$& $\times$  \\ \hline
  \end{tabular}
\caption{Summary of Results. The symbol $\checkmark$ ($\times$) is used when the particular parameter in the column can (can not) be realized within a particular model denoted by the row.}
\label{table:summary}
\end{table}

\section{Results and Conclusion}
\label{sec:conclude}
We have studied the possibility of generating non-zero $\theta_{13}$ by perturbing the $\mu-\tau$ symmetric neutrino mass matrix 
using type II seesaw. The leading order $\mu-\tau$ symmetric mass matrix originating from type I seesaw can be of four different 
types: bi-maximal, tri-bi-maximal, hexagonal and golden ratio mixing, which differ by the solar mixing angle they predict. All these 
four different types of mixing predict $\theta_{23} = 45^o$ and $\theta_{13} = 0$. We use a minimal $\mu-\tau$ symmetry breaking form 
of type II seesaw mass matrix to perturb the type I seesaw mass matrix and determine the strength of type II seesaw term in order to 
generate non-zero $\theta_{13}$ in the correct $3\sigma$ range. We find that except the case of golden ratio mixing with inverted 
hierarchy and $m_3 = 0.001$ eV, all other cases under consideration give rise to correct values of $\theta_{13}$ as can be seen 
from figure \ref{fig:fig1}, \ref{fig:fig2}, \ref{fig:fig3} and \ref{fig:fig4}. We then calculate other neutrino parameters as we 
vary the type II seesaw strength and show their variations as a function of $\sin^2{\theta_{13}}$. We find that bimaximal mixing with 
inverted hierarchy, tri-bimaximal mixing with both normal and inverted hierarchies and hexagonal mixing with normal hierarchy can 
give rise to correct values of neutrino parameters as well as baryon asymmetry. The golden ration mixing is disfavored in our framework 
for both types of neutrino mass hierarchies. 
We have estimated branching ratios for LFV decays like $\mu \rightarrow e + \gamma$ and $\mu \to 3e$ due to the presence of few hundreds 
of GeV mass scale doubly charged scalar triplet Higgs. The estimated value of the branching ratios are found to lie very close to 
the experimental limit. Both prediction of LFV process and origin of non-zero reactor mixing angle are consistent with allowed 
strength of the perturbation term $\omega$ arising from type II seesaw mechanism. 

We summarize our results for all the models under consideration in table \ref{table:summary}. 
We also show the preferred values of Dirac CP phase $\delta$ for successful leptogenesis in table \ref{table:delta}. 
More precise experimental data from neutrino oscillation and cosmology experiments should be able to falsify or verify 
some of the models discussed in this work.  

\section*{Acknowledgments}
Two of the authors, S. Patra and M. K. Das, would like to thank the organizers of the Workshop 
on High Energy Physics and Phenomenology (WHEPP13), held at Puri, Odisha, India during 12-21 
December 2013 where the idea was proposed. The work of Sudhanwa Patra is supported by the Department 
of Science and Technology, Govt. of India under the financial grant SERB/F/482/2014-15. The work of M. K. Das is partially supported by the grant no. 42-790/2013(SR) from UGC, Govt. of India. 

\end{document}